\documentclass[english]{aastex62} 
\usepackage[T1]{fontenc}
\usepackage{textcomp} 
\usepackage{graphicx}
\usepackage{natbib}
\usepackage{textgreek}
\usepackage{amsmath}
\usepackage{color}
\usepackage{babel}

\setcitestyle{yysep={;}}



\begin{document}

\title{Coronal Plasma Characterization via Coordinated Infrared and Extreme Ultraviolet Observations of a Total Solar Eclipse}

\author{Chad A. Madsen}
\affiliation{Center for Astrophysics | Harvard \&{} Smithsonian, 60 Garden St., Cambridge, MA 02138, USA}

\email{cmadsen@cfa.harvard.edu}

\author{Jenna E. Samra}
\affiliation{Center for Astrophysics | Harvard \&{} Smithsonian, 60 Garden St., Cambridge, MA 02138, USA}

\author{Giulio Del Zanna}
\affiliation{Department of Applied Mathematics and Theoretical Physics, CMS, University of Cambridge, Wilberforce Road, Cambridge CB3 0WA, UK}

\author{Edward E. DeLuca}
\affiliation{Center for Astrophysics | Harvard \&{} Smithsonian, 60 Garden St., Cambridge, MA 02138, USA}

\begin{abstract}
We present coordinated coronal observations of the August 21, 2017 total solar eclipse with
the Airborne Infrared Spectrometer (AIR-Spec) and the Extreme-ultraviolet Imaging
Spectrometer (EIS). These instruments provide an unprecedented view of the solar corona in
two disparate wavelength regimes, the near to mid infrared (IR) and the extreme ultraviolet (EUV),
opening new pathways for characterizing the complex coronal plasma environment.  During totality, AIR-Spec sampled coronal IR spectra near the equatorial west limb, detecting strong
sources of Mg VIII, S XI, Si IX, and Si X in two passbands encompassing 1.4 - 4 \textmu{}m. We
apply emission measure (EM) loci analysis to these IR emission lines to test their capacity as coronal temperature diagnostics. The density-sensitive Fe XII
186.9 \AA{}/192.4 \AA{} line pair supplies spatially resolved, line-of-sight
electron densities, supporting the EM loci analysis. From this, we find EM loci
intersections at temperatures of $10^{6.13}$ K at 30 arcsec from the limb and $10^{6.21}$ K at 100
arcsec. Applying the same EM loci analysis to 27 EIS emission lines associated
with seven ion species (Fe X-XIV, S X, and Si X) confirms these results, displaying strong
evidence of isothermal plasma throughout the region. However, the IR EM loci analysis suffers from moderate uncertainties. The likely sources include: poor signal, infrared contamination from a prominence, and photoexcitation by continuum radiation.
Regardless, we demonstrate that EUV spectral data are valuable constraints to coronal infrared emission models, and will be powerful supplements for future
IR solar observatories, particularly DKIST.
\end{abstract}


\section{Introduction}

Total solar eclipses, despite their brevity, can provide rich observational data of coronal emission not otherwise detectable under normal conditions.  Lunar occultation of the solar disk eliminates much of the continuum background scattered by Earth's atmosphere that typically overwhelms optically thin coronal emission in the visible, infrared (IR), and near-visible ultraviolet (UV). 

Eclipse observations of the IR corona began in the mid-20th century when \citet{Blackwell1952} found IR excesses during a 1952 eclipse supporting the existence of the F-corona.  Similar radiometric observations were repeated to both verify models of coronal structure \citep[e.g.][]{Taylor1964} and verify the existence of transient coronal dust originating from solar-cometary interactions \citep{Mizutani1984,Lamy1992,Tollestrup1994,MacQueen1994}. IR spectroscopic observations of the corona during eclipse totality began with \citet{Kurt1962} who sought to improve upon coronagraph measurements under daylight conditions \citep{Firor1962}. Both works provided the first observational confirmations of the forbidden Fe XIII 1.075/1.080 \textmu{}m line pair, the most intense coronal IR lines with respect to the continuum.  This led to several more successful observations of the line pair during later eclipses \citep{Eddy1967,Byard1971,Pasachoff1976,Pasachoff1978,Bao2009,Dima2018}, primarily motivated by the electron density diagnostics associated with their intensity ratio \citep[e.g.][and references therein]{Singh2002}. 

With the advancement of infrared instrumentation came the identification of more coronal IR lines observed during total solar eclipse events \citep[e.g.][]{Olsen1971,Kastner1993,Dima2018}, including Fe XIV 1.27 \textmu{}m, Si X 1.43 \textmu{}m, S XI 1.93 \textmu{}m, Al X 2.75 \textmu{}m, and Mg VIII 3.03 \textmu{}m. Instrumental improvements also drove interest in using infrared lines as probes of coronal magnetic fields \citep[e.g.][]{Judge1998} due in large part to the expected resolvability of emission line splitting due to the Zeeman Effect \citep{Zeeman1897} in the IR. This ultimately provided motivation for construction of the Daniel K. Inouye Solar Telecscope \citep[DKIST,][]{Keil2009,Tritschler2016}, which will provide dedicated ground-based spectroscopic observing in the IR. This investigative approach has the potential to provide a gamut of new physical plasma diagnostics originating from IR emission lines at unprecedented data volumes.  However, diagnostics from IR emission lines are not yet well established.

Unlike the IR regime, physical spectroscopic diagnostics are common tools in the extreme ultraviolet (EUV). The totality of these diagnostic techniques is broad and varied, so we only provide a brief and general history of the subject here. Please see a review by \citet{DelZanna2018b} for a more thorough introduction. Early theoretical work examined the link between collisionally induced atomic transitions and the local electron density environment, especially concerning ion species of Fe \citep{Pottasch1963}.  The development of electron density and temperature diagnostics for plasmas in the EUV accelerated with the groundbreaking spectroheliographic observations by a series of instruments onboard \textit{Skylab} \citep{Reeves1972} such as those used for off-limb density measurments by means of C III line ratios \citep{Doschek1977}. Sounding rocket observations, such as those performed by the Solar Extreme-ultraviolet Rocket Telescope and Spectrograph \citep[SERTS,][]{Neupert1992}, demonstrated the viability of using strong Mg V-IX and Si VIII-X lines in the EUV as both temperature and density diagnostics \citep[e.g.][]{Dwivedi1993,Dwivedi1998}. The advent of dedicated space-based EUV spectroscopy ushered in by instruments such as the Coronal Diagnostics Spectrometer \citep[CDS,][]{Harrison1995} and Solar Ultraviolet Measurements of Emitted Radiation \citep[SUMER,][]{Wilhelm1995} onboard \textit{Solar and Heliospheric Observatory} (SOHO), and the Extreme-ultraviolet Imaging Spectrometer \citep[EIS,][]{Culhane2007} onboard \textit{Hinode} allowed for the refinement of the several diagnostic techniques. EUV spectroscopic diagnostics are now viable for a variety of solar phenomena, including: prominences \citep[e.g.][]{Anzer2007}, coronal holes \citep[e.g.][]{Kashyap2015,Wendeln2018}, coronal mass ejections \citep[e.g.][]{McIntosh2010}, and plasmas not in thermodynamic equilibrium \citep[e.g.][]{Dzifcakova2010,Mackovjak2013}. EUV plasma diagnostics applied to solar spectroscopic data also have application to stellar sources and other astrophysical targets \citep[e.g.][]{MonsignoriFossi1994}. EUV spectroscopic diagnostics are profoundly versatile and robust, providing a benchmark for the expansion of physical diagnostics into other wavelength regimes.

For this work, we draw attention to two varieties of EUV diagnostics.  The first comprises electron density estimates based on intensity ratios of resonant line pairs for a single ion species.  Of particular interest are diagnostics reliant on Fe XII line pairs \citep[e.g.][]{Feldman1981,Feldman1983,Tayal1988,Young2009,Shimizu2017}, which are strong and abundant throughout much of the EUV in both quiescent and active conditions. The second variety consists of temperature diagnostics using emission measure (EM) estimates.  Specifically, we consider an estimate known as the EM loci \citep{Landi2002A,Landi2002B,Landi2012}, which when applied to lines from several distinct ion species determines whether an observed plasma can be approximated as isothermal, and if so, also determines the characteristic plasma temperature associated with that state.

The goals of this work are two-fold: (1) to determine the viability of coronal IR emission lines for use as physical plasma diagnostics, and (2) to demonstrate the value of coordinated EUV observations as a supplement to IR spectroscopic observations. We achieve these goals by analyzing two spectroscopic data sets from coordinated off-limb observations taken during the August 21, 2017 total solar eclipse. The first data set consists of IR spectra from the Airborne Infrared Spectrometer \citep[AIR-Spec,][]{Samra2016} and the second is EUV spectra from EIS. Our general methodology consists of applying EM loci temperature diagnostics to both data sets after using the EUV spectra from EIS for electron density measurements.

This article proceeds as follows: in Section \ref{sec:instobs}, we outline the capabilities of both instruments and describe their respective data sets; Section \ref{sec:dataanls} establishes the theoretical background underpinning our physical diagnostics and details their application to our data sets; in Section \ref{sec:RandD}, we state the results of our diagnostic analysis and discuss their significance; and finally, Section \ref{sec:concl} summarizes the article and extrapolates on related future work.

\section{Instruments \&{} Observations}\label{sec:instobs}

This study focuses on two data sets from observations of the equatorial western limb of the Sun during and shortly after the total solar eclipse of August 21, 2017.  The two data sets originate from two distinct instruments, each of which provides spectroscopic data in different passbands:  AIR-Spec which observes in the IR, and EIS which observes in the EUV.  In this section, we provide instrumental and observational context for those two data sets.

\subsection{IR Spectra from AIR-Spec}

AIR-Spec is a grating spectrometer designed to observe solar eclipses aboard aircraft. Observing at altitudes near 15 km abates the primary tropospheric source of opacity in the near infrared: absorption from dense vibrorotational telluric bands of water vapor.  This permits Earth-based observation of an IR passband ranging between 1 \textmu{}m and 4 \textmu{}m.  The primary goal of AIR-Spec is to observe magnetically sensitive emission lines in this passband, which could pave the way for direct measurements of the coronal magnetic field in future investigations.  However, AIR-Spec also provides a unique opportunity to test the diagnostic potential of coronal IR lines, which we establish as the primary goal of this study.

The spectroscopic capabilities of AIR-Spec provide a reliable means of achieving this goal. Its setup features a nonrotating slit approximately 1.5 $R_{\odot}$ in projected length, suitable for coverage of extended coronal emission.  A single InSb detector records the incoming spectra in two channels comprising overlapping first-order and second-order spectra.  In first-order, the passbands cover ranges from 2.83 \textmu{}m to 3.07 \textmu{}m and 3.75 \textmu{}m to 3.98 \textmu{}m with a spectral sampling of 2.4 \AA{} pixel$^{\text{-}1}$.  In second order, the spectral sampling improves by a factor of two, setting the wavelength bounds at half the first-order values: 1.42 \textmu{}m to 1.54 \textmu{}m and 1.87 \textmu{}m to 1.99 \textmu{}m. The spectral resolution is about 15 \AA\ in first order and 7.5 \AA\ in second order. The spatial resolution ranges from 11--13 arcsec, and the projected spatial pixel size is 2.3 arcsec.

The passbands allow for the observation of emission from five forbidden magnetic dipole transitions in the fine structure of heavy metallic ions. These five lines are: Si X 1.43 \textmu{}m, S XI 1.92 \textmu{}m, Fe IX 2.84 \textmu{}m, Mg VIII 3.03 \textmu{}m, and Si IX 3.94 \textmu{}m.  The measured line wavelengths and associated transitions are summarized in Table \ref{tab:AIRSpecEMLociLines}.  
\begin{table}[!hbtp]
\begin{center}
\begin{tabular}{c c l l}
\hline \hline \\[-3mm]
Ion & Wavelength [\textmu{}m] & Lower State & Upper State \\
\hline \\[-3mm]
Si X & 1.431 & $2s^{2}2p$ $^{2}P_{1/2}$ & $2s^{2}2p$ $^{2}P_{3/2}$ \\
S XI & 1.921 & $2s^{2}2p^{2}$ $^{3}P_{0}$ & $2s^{2}2p^{2}$ $^{3}P_{1}$ \\
Mg VIII & 3.028 & $2s^{2}2p$ $^{2}P_{1/2}$ & $2s^{2}2p$ $^{2}P_{3/2}$ \\
Si IX & 3.935 & $2s^{2}2p^{2}$ $^{3}P_{0}$ & $2s^{2}2p^{2}$ $^{3}P_{1}$ \\[1mm]
\hline \hline \\[-3mm]
\end{tabular}
\caption{List of AIR-Spec's Target IR Lines.}
\label{tab:AIRSpecEMLociLines}
\end{center}
\end{table}
All of the lines originate from ground-state fine structure transitions induced by electron collisions.  Fine structure transitions are expected to occur far more often than resonance transitions starting at large principal quantum numbers, making these four lines a likely source of bright IR emission in the quiet-Sun corona \citep{Judge1998,DelZanna2018}.

\begin{figure}[!hbtp]
\begin{center}
\includegraphics[trim=1cm 3cm 1cm 3cm,clip=true,width=\linewidth]{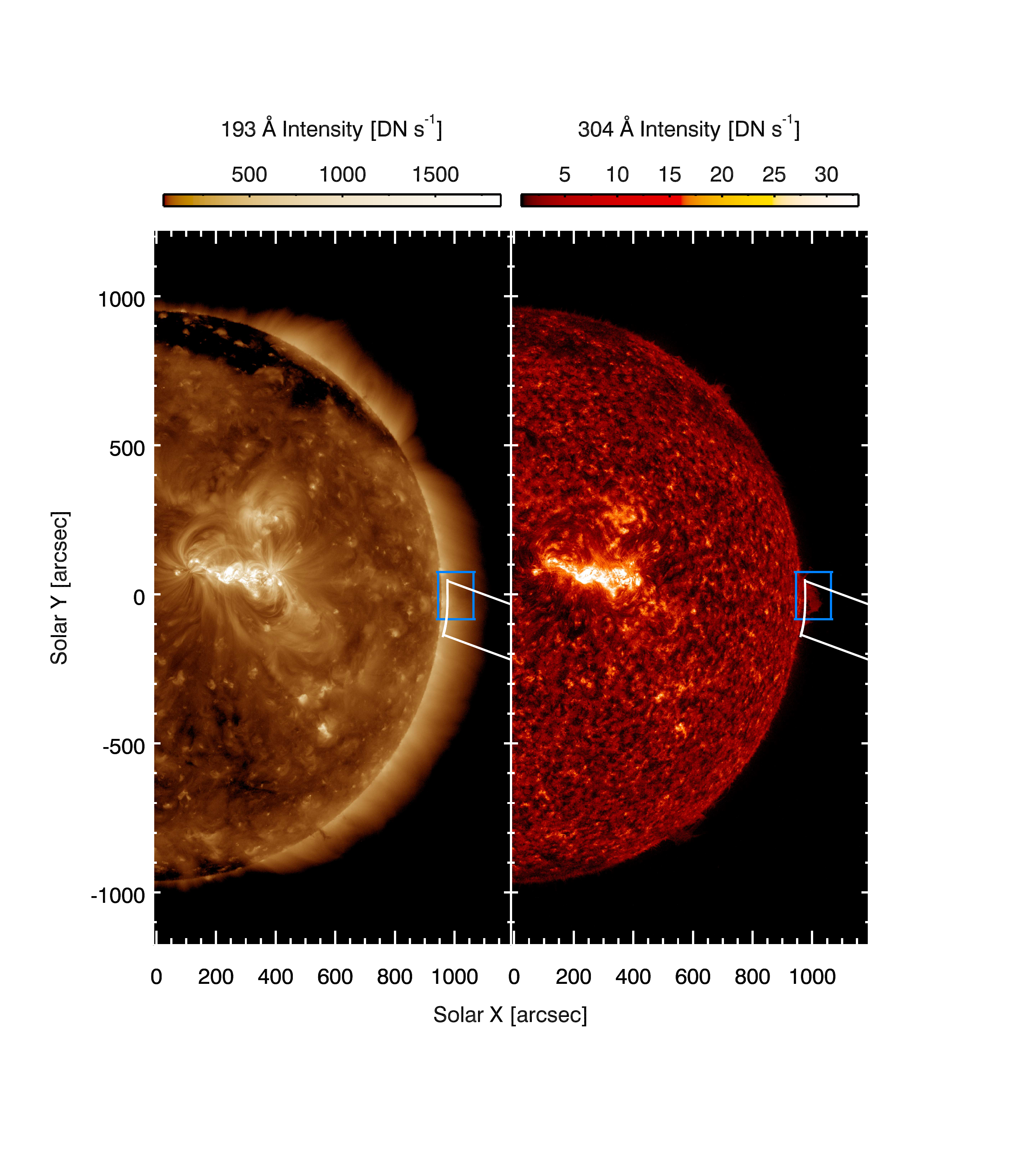}
\caption{Half-disk AIA 193 \AA{} (left) and AIA 304 \AA{} (right) context images for August 21, 2017 at approximately 21:37 UT depicting the AIR-Spec (white) and EIS (blue) fields-of-view on the western limb.}
\label{fig:AIAfullcon}
\end{center}
\end{figure}

AIR-Spec's first eclipse observation took place on August 21, 2017 from 14.3 km over western Kentucky, near the location of maximum totality duration.  From the National Science Foundation/National Center for Atmospheric Research (NSF/NCAR) Gulfstream V, AIR-Spec observed totality for about four minutes, 1.5 minutes longer than an observer directly below at ground level.  During this time, the slit sampled coronal emission from four different pointings.  The second of these pointings (henceforth referred to as \textit{pointing \#{}2}) provided the IR spectral data used in this work.  It sampled about 41 s of data from a region near the equatorial western limb occupied by a small prominence and quiet-Sun coronal plasma. The effective field-of-view (FOV) for pointing \#{}2 is shown on half-disk \textit{Solar Dynamics Observatory}/Atmospheric Imaging Assembly \citep[SDO/AIA,][]{Lemen2012} images in Figure \ref{fig:AIAfullcon}.  
Here, we see the slit rotated $21.2^{\circ}$ clockwise from the east-west horizontal. The eastern end of the FOV is a roughly circular arc resting about 25 arcsec above the limb; this boundary represents the slit's intersection with the occulting lunar disk.  During the 2017 flight, AIR-Spec used an open-loop image stabilization system to compensate for most of the aircraft's motion and prevent smearing within an exposure time.  Some frame-to-frame jitter remained, broadening the sampling range perpendicular to the slit and effectively expanding the FOV's north-south width to about 200 arcsec.

\begin{figure}[!hbtp]
\begin{center}
\includegraphics[width=0.6\linewidth]{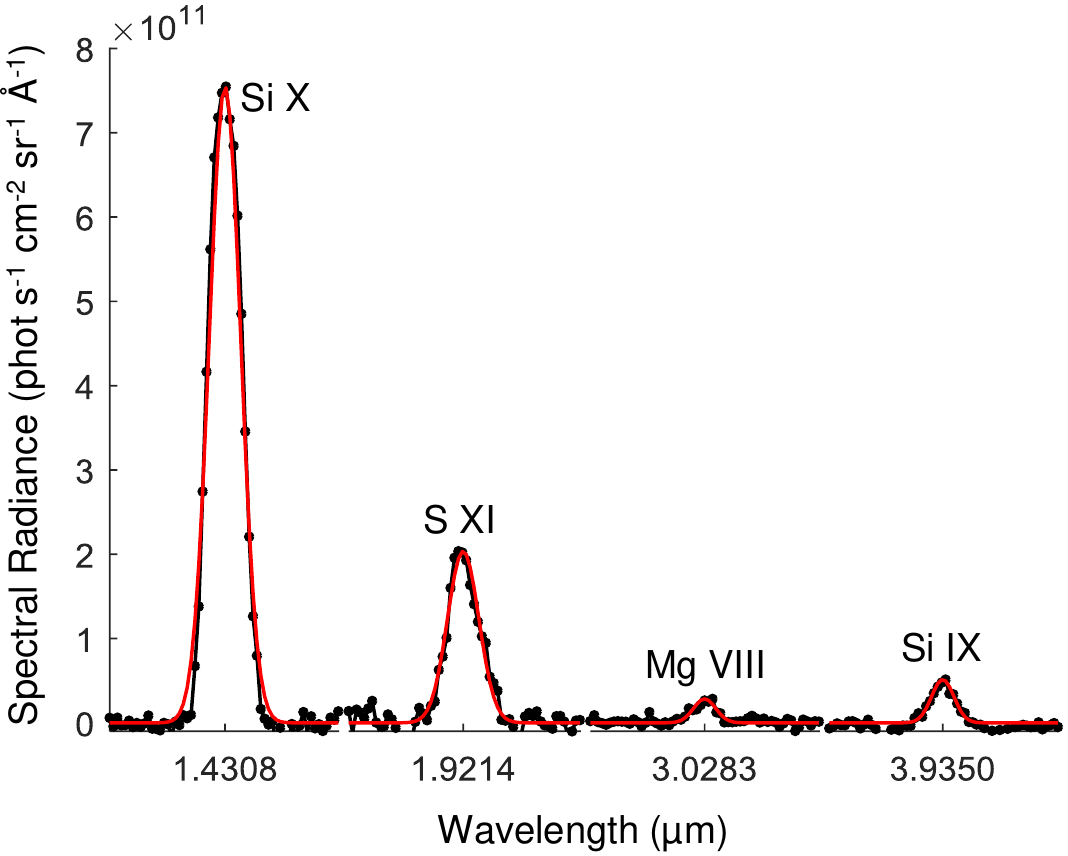}
\caption{Temporally and spatially averaged AIR-Spec spectral radiance (black dots) at pointing \#{}2 over 35 arcsec near the lunar limb. Gaussian fits to the lines are shown in red. Wavelengths are reported in vacuum.}
\label{fig:airspec-lines}
\end{center}
\end{figure}

The data recorded on the day of the eclipse underwent a routine series of calibrations. The radiometric throughput was calibrated in first and second order by exposing the spectrograph to the solar disk in the presence of neutral density ($T=10^{-3})$ and order-isolating filters. A linear wavelength calibration was performed using the photospheric absorption spectrum to estimate the slope in \AA/pixel and hydrogen emission lines from the prominence to estimate the wavelength of the first spectral pixel at the time of the observation. To reduce the thermal background and dark current, the spectrometer was passively cooled to below 150 K using liquid nitrogen, a closed-cycle chiller cooled the infrared detector to 59 K, and the interface between the spectrometer and IR camera was packed in solid carbon dioxide to reduce its thermal emission onto the detector. The significant residual thermal background, nonuniform in space and nonlinear in time, limited the exposure time to 60 ms and required empirical modeling to remove.  For this reason, only the strong first Paschen line (1.88 \textmu{}m) from the prominence appeared in the individual frames.  However, coadding all 622 frames produced statistically significant detections for all four predicted lines. Figure \ref{fig:airspec-lines} shows the four line profiles measured during pointing \#{}2, averaged in time and over the 35 arcsec nearest the lunar limb.  Additionally, AIR-Spec also detected two weak lines: one centered at 2.843 \textmu{}m and another centered at 2.853 \textmu{}m in air. The former line was originally identified as an Fe IX emission line predicted by \citet{Judge1998} and \citet{DelZanna2018}; however, it was later discovered to be a ghost image of Si X 1.43 \textmu{}m produced by the uncoated side of the slit-jaw \citep[][erratum]{Samra2018}. The 2.853 \textmu{}m line was determined to be genuine. Its identity remains uncertain, although it stands as a strong candidate for the predicted Fe IX line.  Given its questionable identity and weak signal, we did not include this line in our analysis.  

\subsection{EUV Spectra from EIS}

\begin{figure}[!hbtp]
\begin{center}
\includegraphics[trim=1cm 0cm 1cm 0cm,clip=true,width=\linewidth]{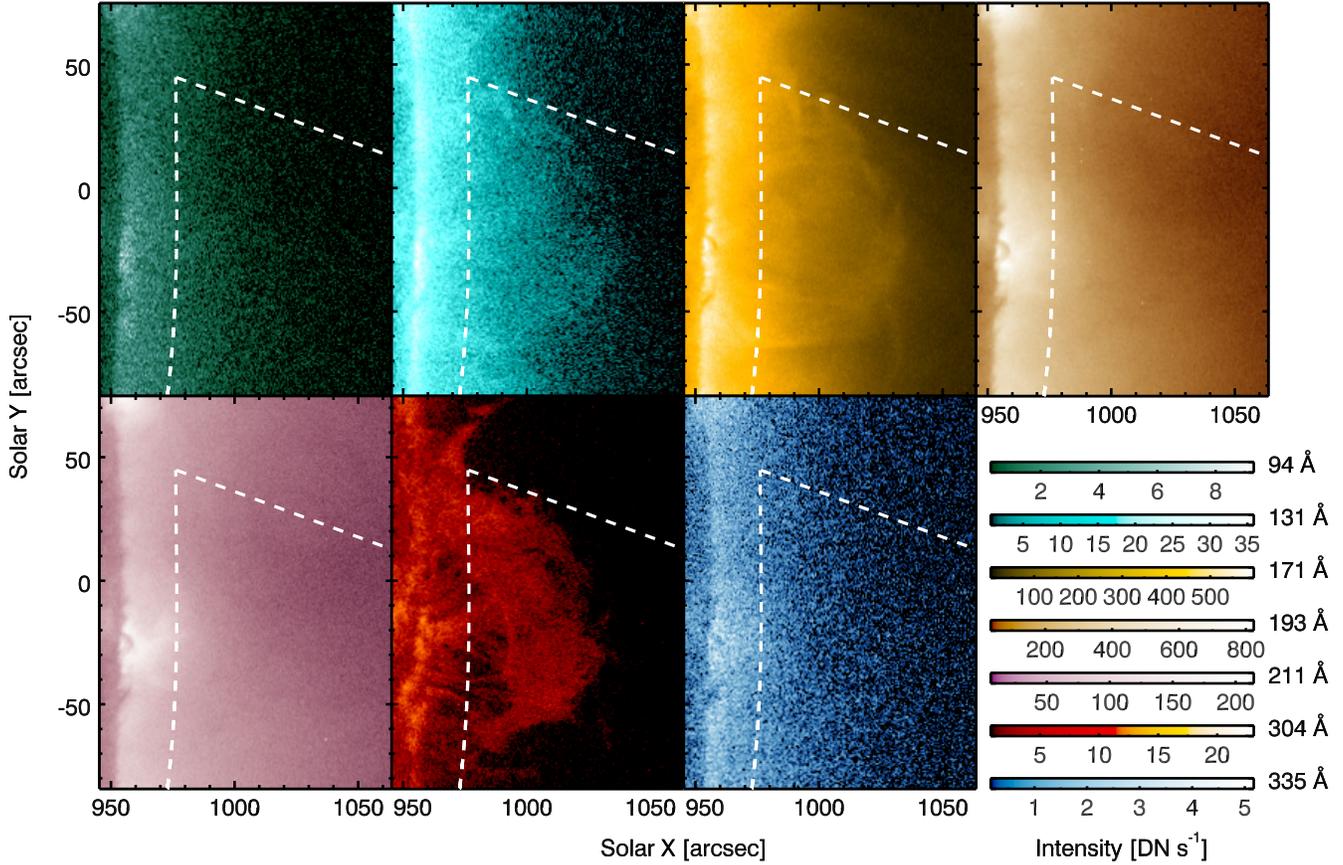}
\caption{AIA cutouts of the EIS field-of-view across all seven AIA EUV channels taken on August 21, 2017 near 21:37 UT. The white dashed line indicates the northeastern boundary of the AIR-Spec field-of-view.}
\label{fig:AIAcutcon}
\end{center}
\end{figure}

Coordinated observations with EIS onboard \textit{Hinode} took place approximately two hours after the AIR-Spec observations, starting around 21:36 UT and ending about two hours later at 23:36 UT.  EIS pointed approximately 50 arcsec above the equatorial western limb, sampling a single west-to-east, 60-step raster with a step size of 2 arcsec and a step exposure time of 120 s.  Since the vertical pointing of EIS drifts as a function of wavelength, we establish 186.8 \AA{} as our pointing and FOV reference.  After correcting for spacecraft jitter and coaligning the coordinate system to that of AIA, we find the central pointing at 186.8 \AA{} to be $\left(X\text{,}Y\right)=\left(1005.5\text{",-}4.1\text{"}\right)$ and the slit raster coverage to be $117.9\text{"}\times{}159.0\text{"}$ spanning from $\left(945.6\text{",-}84.1\text{"}\right)$ at the southeasternmost corner to $\left(1063.6\text{",}74.9\text{"}\right)$ at the northwesternmost corner as shown in Figure \ref{fig:AIAfullcon}. Approximately 80\% of the EIS FOV area overlaps with the northeasternmost portion of the effective AIR-Spec FOV. AIA cutouts of the EIS FOV are seen in Figure \ref{fig:AIAcutcon}.  
AIA 131 \AA{}, 171 \AA{}, and 304 \AA{} all show evidence of a small prominence extending about 50 arcsec from the limb.  Additionally, AIA 193 \AA{}, 211 \AA{}, and to a much lesser extent, 94 \AA{} and 335 \AA{} show two elliptical regions of enhanced intensity rising about 10 arcsec off of the limb, one of which is found near $Y=\text{-}30$ arcsec and other occupying the northern limit of the FOV. Overall, the plasma environment appears to be typical of quiet-Sun conditions.

Prioritizing throughput over resolution guided our choice of observing parameters.  We used the wider 2" slit, as opposed to the 1" slit, for precisely this purpose.  To provide the most detailed diagnostics possible, we used a full readout of EIS's CCD.  Consequently, this required a dramatic shrinking of the vertical component of the FOV to accommodate EIS's telemetry limitations.  The full-readout mode consists of four broad wavelength windows: two in the short wavelength regime (166.4 \AA{} $\to$ 189.2 \AA{} and 189.2 \AA{} $\to$ 212.0 \AA{}), and two in the long wavelength regime (245.9 \AA{} $\to$ 268.7 \AA{} and 268.7 \AA{} $\to$ 291.4 \AA{}.)

Before analysis, we performed a series of calibrations to transform the level-0 EIS spectral data into level-1 science-ready data, which comprised the following: dark current and bias subtraction, removal of cosmic rays \citep{Pike2000}, flagging and removal of hot and warm pixels, an orbital wavelength correction, and a radiometric calibration which converts units of spectral radiance from digital numbers (DN) to erg s$^{\text{-}1}$ cm$^{\text{-}2}$ sr$^{\text{-}1}$ \AA{}$^{\text{-}1}$.  For the radiometric calibration, we applied a time-dependent sensitivity decay model \citep{Warren2014} to the original pre-launch calibration \citep{Lang2006}.  The missing pixel values resulting from cosmic ray and warm/hot pixel removal were replaced using a robust hierarchical interpolation scheme that minimized the influence of adjacent missing pixels.   Interpolated values were derived from neighboring pixels within the same data column, which helps preserve the shape of spectral line profiles.  The algorithm replaces an isolated missing pixel value with the average of the pixel values above and below it.  However, if one of those neighboring pixel values is missing, then the algorithm will try to find a suitable replacement by taking a weighted average of the nonmissing pixel value and the second-nearest neighbor (or third-nearest neighbor if that too is missing) on the opposite side. If both neighboring pixel values are missing, then the algorithm takes the average of the second-nearest neighbors on either side.  If all of these scenarios fail, then the algorithm replaces the missing pixel value with the value assigned to one of it's formally missing neighbors.  This process was iterated until all missing values were accounted for.

Fourteen frames were rejected from the analysis due to various reasons.  Five of these coincided with the spacecraft's crossing of the south Atlantic Anomaly.  Another four were associated with partial or complete dropouts of the CCD readout.  The remaining five were associated with an apparent uniform drop in intensity across the detector that could not be explained by accompanying changes to the metadata.  For the remainder of this paper, we refer to this incident as the \textit{unexplained dimming}.  We also removed a defective row of pixels in the top third of the CCD readout which exhibited anomalously high pixel values.

\section{Data Analysis}\label{sec:dataanls}

Analysis of the two data sets comprises two distinct steps.  First, we use the EIS spectra to produce resolved electron density estimates throughout the EIS FOV.  Second, we use the electron density estimates to calculate EM loci for several ion species found in the EIS and AIR-Spec data.  The intent of the EM loci analysis is two-fold: (1) to determine if the coronal plasma in the EIS FOV is characteristically isothermal, and if so, (2) to determine the spatial distribution of the associated isothermal temperatures.

\subsection{Electron Density Analysis}\label{subsec:elecdense}

We determine electron densities using well-established diagnostics applied to resonant Fe XII emission lines \citep[e.g.][]{Pottasch1963,Feldman1981,Feldman1983,Tayal1988,Young2009,Shimizu2017} found in the EIS spectral window. In particular, we focus on three lines associated with transitions from the $3s^{2}3p^{2}3d$ configuration to the $3s^{2}3p^{3}$ configuration: 186.85 \AA{} (${}^{2}F_{5/2}\to{}{^2}D_{3/2}$), 186.89 \AA{} (${}^{2}F_{7/2}\to{}{}^{2}D_{5/2}$), and 192.39 \AA{} (${}^{4}P_{1/2}\to{}^{4}S_{3/2}$). Information about these lines are summarized in Table \ref{tab:DenseSenseLines}.
\begin{table}[!hbtp]
\begin{center}
\begin{tabular}{c l l c}
\hline \hline \\[-3mm]
Wavelength [\AA] & Lower State & Upper State & $A_{lu}$ [s$^{\text{-}1}$] \\[1mm]
\hline \\[-3mm]
186.854 & $3s^{2}3p^{3}$ $^{2}D_{3/2}$ & $3s^{2}3p^{2}3d$ $^{2}F_{5/2}$ & $1.010 \times 10^{11}$ \\
186.887 & $3s^{2}3p^{3}$ $^{2}D_{5/2}$ & $3s^{2}3p^{2}3d$ $^{2}F_{7/2}$ & $1.080 \times 10^{11}$\\
192.394 & $3s^{2}3p^{3}$ $^{4}S_{3/2}$ & $3s^{2}3p^{2}3d$ $^{4}P_{1/2}$ & $8.830 \times 10^{10}$\\[1mm]
\hline \hline \\
\end{tabular}
\caption{List of EIS EUV Fe XII Lines Used for Electron Density Analysis}
\label{tab:DenseSenseLines}
\end{center}
\end{table}
The former two lines comprise a self-blend which we treat as a single line called Fe XII 186.85/.89 \AA{} for the purposes of this analysis.  The diagnostic method used in this work concerns a relationship between intensity ratios of Fe XII emission lines and the electron densities associated with their line-of-sight source volumes.  It works by exploiting the efficiency difference by which collisional processes populate the upper states of the transitions associated with each line.

We proceed by establishing the basic framework of this diagnostic. For a more detailed review, please refer to \citep{DelZanna2018b}. To begin, we find the intensity $I$ at wavelength $\lambda$ for emission due to a transition from an upper state $j$ to a lower state $i$ within species $X$ at ionization state $+m$ assuming an optically thin plasma:
\begin{equation}\label{eq:intens}
I_{ij}\left(X^{+m};\lambda_{ij}\right)=\frac{1}{4\pi{}R^{2}}\int_{V}\epsilon_{ij}\left(X^{+m}\right)dV=\frac{h\nu_{ij}}{4\pi{}R^{2}}\int_{V}n_{j}\left(X^{+m}\right)A_{ij}\left(X^{+m}\right)dV
\end{equation} 
where $R$ is the distance from the emission source to the observer, $V$ is the line-of-sight volume of the emission source, $\epsilon_{ij}\left(X^{+m}\right)$ is the volume emissivity for a transition between states $i$ and $j$ of ion species $X$ at ionization state $+m$, $\nu_{ij}$ is the frequency associated with $\lambda_{ij}$, $n_{j}\left(X^{+m}\right)$ is the volume number density of ions of species $X$ in ionization state $+m$ at upper state $j$ within the source volume, and $A_{ij}\left(X^{+m}\right)$ is the Einstein coefficient for spontaneous emission of transition $j\to{}i$ for the same ion species \citep{Einstein1916}.  Note that our definition of intensity, Eq. (\ref{eq:intens}), uses energy units instead of photon units.

\begin{figure}[!hbtp]
\begin{center}
\includegraphics[trim=2cm 0cm 0cm 0cm,clip=true,width=\linewidth]{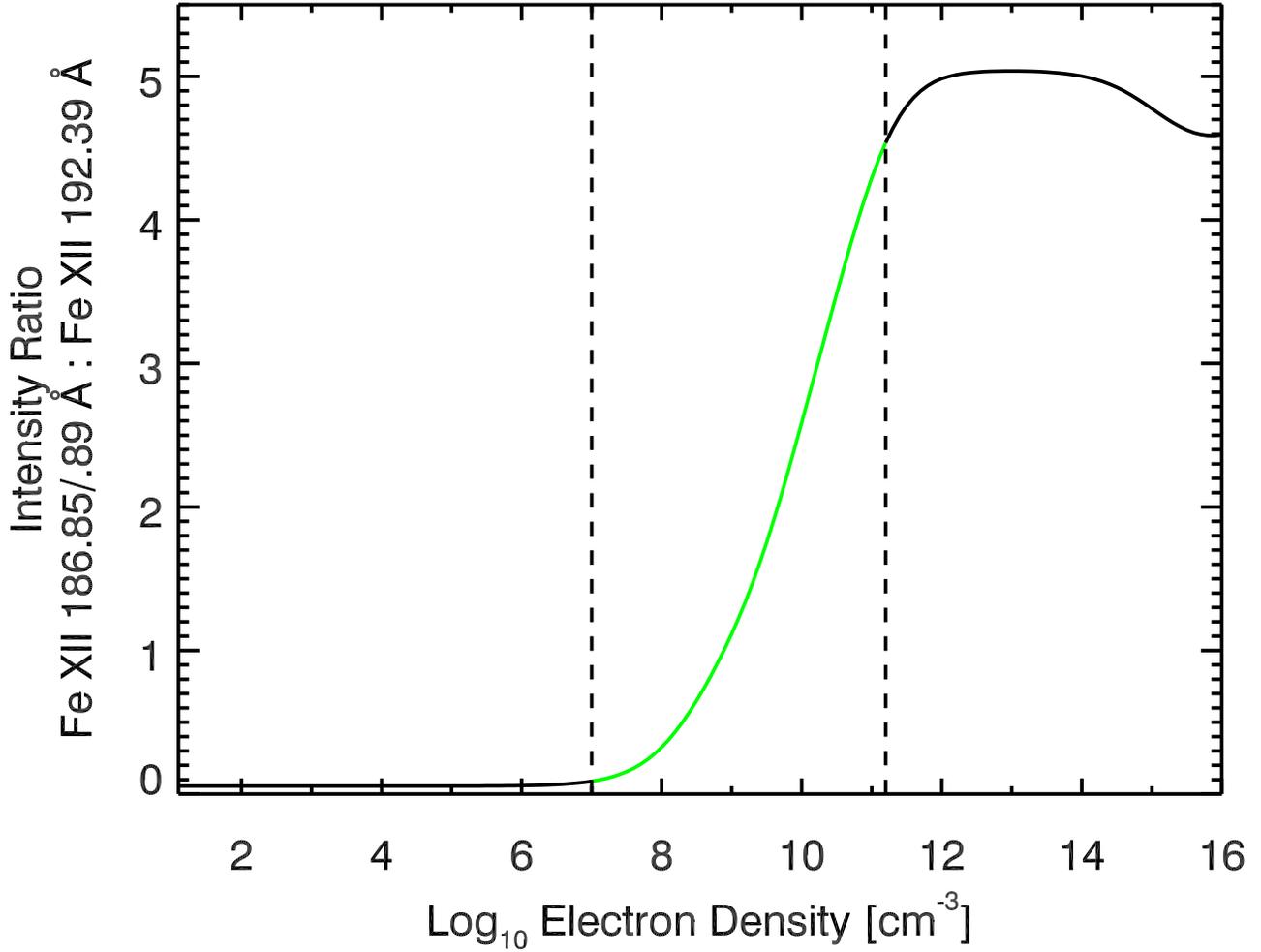}
\caption{Plot of intensity ratio between Fe XII 186.85/.89 \AA{} and Fe XII 192.39 \AA{} versus electron density.  The dashed lines indicate the lower and upper bounds of the region where a one-to-one relationship between the electron density and the intensity ratio is guaranteed, as indicated by the green section of the plot.}
\label{fig:densrat}
\end{center}
\end{figure}

$A_{ij}\left(X^{+m}\right)$ only depends on the upper and lower states of the transition, so all intensity variation with respect to electron density originates from $n_{j}\left(X^{+m}\right)$.  However, calculating $n_{j}\left(X^{+m}\right)$ is no simple task since we need information on all possible transitions to and from the upper state $j$ within the ion.  If we assume that the plasma is in ionization equilibrium and the timescale for electron collisions is much smaller than the observation time, then we can determine $n_{j}\left(X^{+m}\right)$ using the principle of detailed balance:
\begin{equation}\label{eq:detbalance}
n_{j}\left(X^{+m}\right)\left[n_{e}\displaystyle\sum_{k\neq{}j}C_{kj}\left(X^{+m}\right)+\displaystyle\sum_{k<j}A_{kj}\left(X^{+m}\right)\right] =n_{e}\displaystyle\sum_{k\neq{}j}n_{k}\left(X^{+m}\right)C_{jk}\left(X^{+m}\right)+\displaystyle\sum_{k>j}n_{k}\left(X^{+m}\right)A_{jk}\left(X^{+m}\right)
\end{equation}
where $n_{e}$ is the electron number density and $C_{qp}\left(X^{+m}\right)$ is the rate coefficient for a collisionally induced transition (either excitation or de-excitation) from a state $p$ to some other state $q$ in ion species $X^{+m}$.  According to eq. (\ref{eq:detbalance}), solving for $n_{j}\left(X^{+m}\right)$ requires knowledge of the number densities for every other state in the ion. Assuming there are $N_{s}$ non-negligible states in the ion, we can apply eq. (\ref{eq:detbalance}) to each of those states, producing a system of $N_{s}$ linear equations. To make this set of equations inhomogeneous and thus solvable for $n_{j}\left(X^{+m}\right)$, we introduce:
\begin{equation}
\displaystyle\sum_{k}n_{k}\left(X^{+m}\right)=n\left(X^{+m}\right)
\end{equation}
where $n\left(X^{+m}\right)$ is the total volume number density of ion $X^{+m}$ in the source volume, a value that depends on the yet unknown temperature of the plasma.  Fortunately, $n\left(X^{+m}\right)$ cancels when $n_{j}\left(X^{+m}\right)$ is taken in ratio with another state density. Coupling this with the assumption that the temperature and density are uniform and constant within the source volume, we can then use eqs. (\ref{eq:intens}) and (\ref{eq:detbalance}) to relate the intensity ratio of two emission lines to the electron density. The functional form of this relationship is that of a ratio of two higher-order polynomials of the electron density, the coefficients of which are composed of $A$ and $C$ values in elementary combinations.  The complexity of this relationship is dependent entirely upon the choice of line pair.  While any two lines originating from the same species will satisfy our assumptions, only a rare few will provide a simple enough relationship between their intensity ratio and the electron density to be useful for our analysis. In general, we need a line pair that provides the following: (1) a one-to-one relationship between the intensity ratio and the electron density within the domain of typical coronal densities, and (2) a wide enough range of intensity ratios within that same domain to provide precise electron density estimates within instrumental uncertainties. By using atomic data provided by CHIANTI version 7 \citep{Landi2012}, we find that the Fe XII line pair, 186.85/.89 \AA{} and 192.39 \AA{}, satisfies both of these properties as seen in Figure \ref{fig:densrat}. The intensity ratio range is broad and corresponds to typical quiet-Sun coronal electron density values, spanning from $10^{7}$ cm$^{\text{-}3}$ to about $10^{11}$ cm$^{\text{-}3}$ where the one-to-one relationship breaks just prior to a local maximum. Physically, Figure \ref{fig:densrat} shows a transitions between two atomic equilibrium states.  At low electron densities, the plasma is defined by the absence of collisions where level populations are determined entirely by spontaneous transitions.  On the contrary, at high electron densities, electron collisions become so overwhelming that collisional de-excitation rivals spontaneous emission, producing a stable state for level populations. 

\begin{figure}[!hbtp]
\begin{center}
\includegraphics[width=\linewidth]{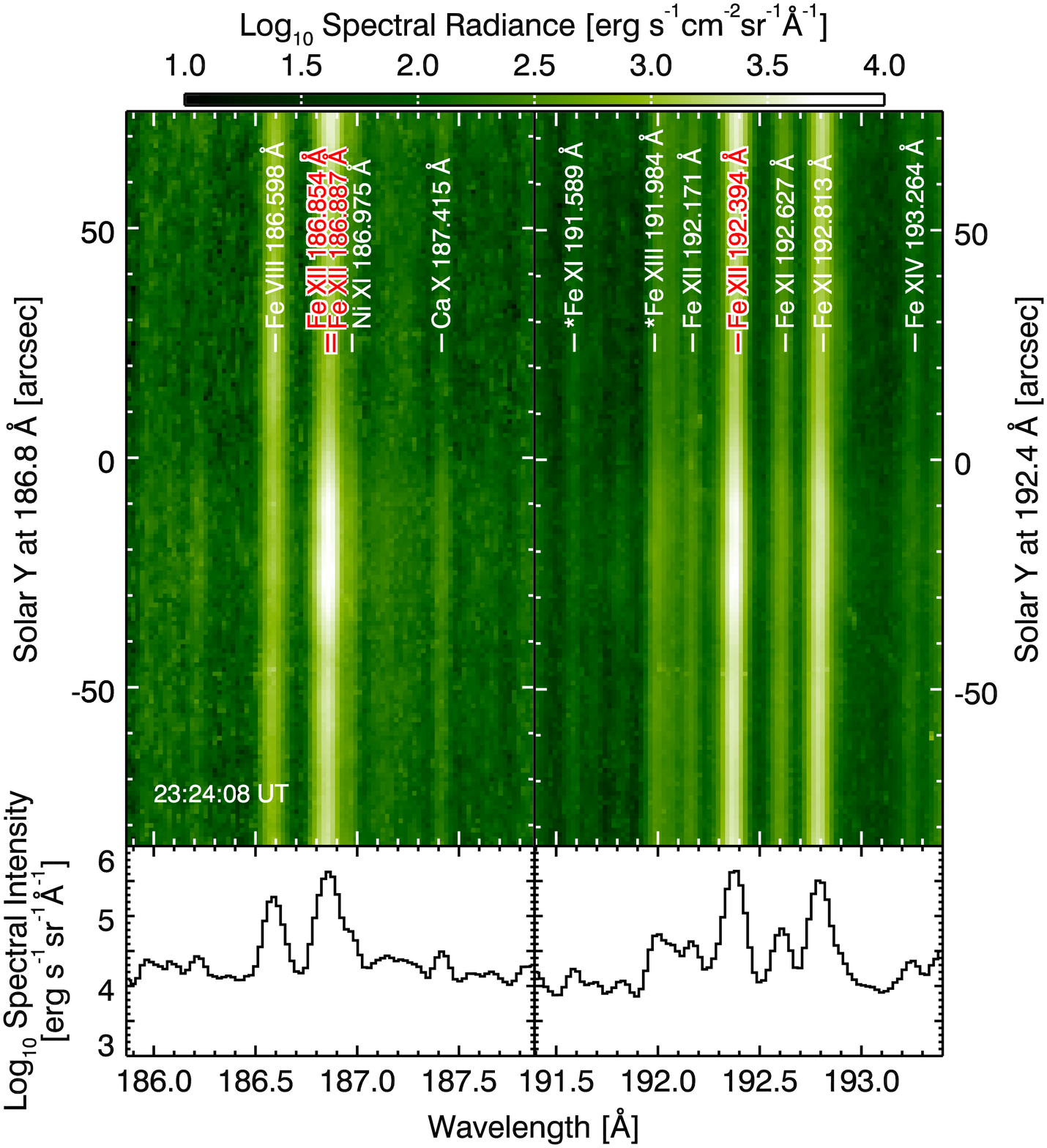}
\caption{Two spectroscopic windows from the EIS detector centered on Fe XII 186.85/.89 \AA{} (upper left panel) and 192.39 \AA{} (upper right panel) from a full-CCD exposure taken on August 21, 2017 at 23:24:08 UT.  These lines are indicated by red labels while nearby lines are labeled in white. Labels preceded by an asterisks (*) indicate theoretical transitions predicted by CHIANTI version 7 \citep{Landi2012}. The lower two panels depict the columnar sums of the spectral radiance, producing plots of the spectral intensity.  Note that both windows span slightly different solar Y ranges.}
\label{fig:linecon}
\end{center}
\end{figure}

The line pair is also suitable for precise intensity measurements.  Examples of the 186.85/.89 \AA{} and 192.39 \AA{} lines are shown in Figure \ref{fig:linecon}. 
Here we find strong, but not uniform, signal throughout the vertical profiles of the lines, approaching spectral radiances on the order of $10^{4}$ erg$^{\text{-}1}$s$^{\text{-}1}$cm$^{\text{-}2}$sr$^{\text{-}1}$\AA$^{\text{-}1}$.  Since the dominant broadening mechanism in the EUV is Doppler broadening, the lines display markedly Gaussian profiles suitable for least-squares fitting.  However, the fitting process is only reliable if both lines are unblended with lines from other ion species.  This is not a problem for Fe XII 192.39 \AA{}, but the profiles of 186.85/.89 do overlap the nearby Ni XI 186.96 \AA{}.  However, Ni XI 186.96 \AA{} is over one order of magnitude less intense at its peak than 186.85/.89.  Furthermore, their centers are separated enough to allow us to apply a multiple-Gaussian fit (MGF) model.  We apply MGFs to each row of both 2 \AA{}-wide windows centered on each line of the pair, corresponding to the horizontal ranges seen in Figure \ref{fig:linecon}. Before we apply the fitting routine, we coalign the FOV of the 192.39 \AA{} to that of 186.85/.89 \AA{} by shifting the 192.4 \AA{} window up 0.44 arcsec and interpolating accordingly.  The MGF model features three parameters applied to each line \textemdash{} peak intensity, line width, Doppler shift \textemdash{} and two background parameters \textemdash{} a constant component and a linear component \textemdash{} per window.  We then determine the integrated intensity of each line using the post-fit peak intensity and line width values, take their ratio, and then apply it to the intensity ratio vs. electron density model shown in Figure \ref{fig:densrat}.

\subsection{Temperature Diagnostics via EM Loci Analysis}\label{subsec:tempdiag}

With electron densities available, we can now apply temperature diagnostics to both the AIR-Spec and EIS data sets.  For this, we use emission measure (EM) loci analysis, which we describe similarly to \citet{Landi2002A,Landi2002B,Landi2012}. First, we express the intensity, Eq. (\ref{eq:intens}), in a more useful form:
\begin{equation}
I_{ij}\left(X^{+m};\lambda_{ij}\right)=\frac{1}{4\pi{}R^{2}}\int_{V}G_{ij}\left(X^{+m};T,n_{e}\right)n_{e}^{2}dV
\end{equation}
where $G\left(X^{+m};T,n_{e}\right)$ is the contribution function associated with transition $j\to{}i$ in ion $X^{+m}$, defined as:
\begin{equation}
G\left(X^{+m};T,n_{e}\right)\equiv{}\frac{h\nu_{ij}n_{j}\left(X^{+m}\right)A_{ij}\left(X^{+m}\right)}{n_{e}^{2}}
\end{equation}
and $T$ is the plasma temperature of the emission source volume. With electron densities known, the temperature dependence originates solely from $n_{j}\left(X^{+m}\right)$ via the total ion species number density, $n\left(X^{+m}\right)$.  
The contribution function is useful in this scenario since it allows us the consolidate all the the species-dependent quantities under one value.  This becomes important when we assume that the electron density and plasma temperature are uniform and constant values, $n_{e0}$ and $T_{0}$, respectively, in the source volume:
\begin{equation}\label{eq:constTN}
I_{ij}\left(X^{+m};\lambda_{ij}\right)=\frac{G_{ij}\left(X^{+m};T_{0},n_{e0}\right)}{4\pi{}R^{2}}\int_{V}n_{e0}^{2}dV\equiv\frac{G_{ij}\left(X^{+m};T_{0},n_{e0}\right)}{4\pi{}R^{2}}EM
\end{equation}
where we define the emission measure, $EM\equiv\int_{V}n_{e}^{2}dV$, a quantity that does not depend on species-specific properties.  We then rearrange Eq. (\ref{eq:constTN}) as follows:
\begin{equation}\label{eq:reconstTN}
EM=\frac{4\pi{}R^{2}I_{ij}\left(X^{+m};\lambda_{ij}\right)}{G_{ij}\left(X^{+m};T_{0},n_{e0}\right)}
\end{equation}
Since the emission measure is the same regardless of ion species or transition, the right-hand side of Eq. (\ref{eq:reconstTN}) must be as well.  This leads us to define the following function:
\begin{equation}\label{eq:eml}
EML\left(T\right)\equiv\frac{I_{ij}\left(X^{+m};\lambda_{ij}\right)}{G_{ij}\left(X^{+m};T,n_{e0}\right)}
\end{equation}
which we name the \textit{emission measure loci} or simply \textit{EM loci}. From here, we can establish the diagnostic: If the temperature and electron density are uniform and constant within a given source volume, then there exists an isothermal temperature value, $T_{0}$, such that $EML\left(T_{0}\right)$ is the same for all transitions and all ion species. In practice, this means we can use the measured values of $n_{e0}$ from the process outlined in Subsection \ref{subsec:elecdense} to calculate contribution functions within a broad temperature range, and then using our measured values of $I_{ij}$, we can determine the EM loci for each emission line detected in the EIS and AIR-Spec data sets.  We can then check if all the EM loci curves intersect at a common temperature value.  If so, then the plasma in the source volume can be approximated as isothermal with characteristic temperature $T_{0}$. For the purposes of this work, we establish a modified definition of the EM loci we denote as $EML^{*}$ which holds the same diagnostic properties as $EML$ but instead consolidates all observed lines of a single ion species:
\begin{equation}
EML^{*}\left(T\right)\equiv\sum_{ij\in{}X^{+m}}I_{ij}\left(X^{+m};\lambda_{ij}\right)/\sum_{ij\in{}X^{+m}}G_{ij}\left(X^{+m};T,n_{e0}\right)
\end{equation}
which allows for greater effective signal for each instance of the EM loci function.  We calculate contribution functions in terms of temperature for each transition using the electron density values measured in Subsection \ref{subsec:elecdense} and solving for $n_{j}\left(X^{+m}\right)$ in Eq. (\ref{eq:detbalance}) using a model of solar elemental abundances \citep{Asplund2009}, and a model of ionization equilibrium and other atomic values from CHIANTI version 7.

We apply EM loci analysis to both the EIS and AIR-Spec data sets.  The EIS data set provides a robust measurement of the isothermal temperature which we can use to verify the effectiveness of the analysis as applied to the AIR-Spec IR spectra. We begin analysis on the EIS data by identifying as many suitable emission lines as possible in the full-CCD readout.  We only select lines that satisfy the following properties: (1) the lines must be detected up to $5\sigma$ at all raster positions, and (2) the lines must be free of all unresolved blends with lines from other ion species.  Twenty-seven lines satisfy these properties, representing seven distinct ions species: Si X, S X, and Fe X-XIV. All lines and their associated transitions are listed in Table \ref{tab:EISEMLociLines}.
\begin{table}[!hbtp]
\begin{center}
\begin{tabular}{c c l l}
\hline \hline \\[-3mm]
Ion & Wavelength [\AA] & Lower State & Upper State \\[1mm]
\hline \\[-3mm]
Si X & 253.790 & $2s^{2}2p$ $^{2}P_{1/2}$ & $2s2p^{2}$ $^{2}P_{3/2}$ \\
	& 258.374 & $2s^{2}2p$ $^{2}P_{3/2}$ & $2s2p^{2}$ $^{2}P_{3/2}$ \\
	& 261.056 & $2s^{2}2p$ $^{2}P_{3/2}$ & $2s2p^{2}$ $^{2}P_{1/2}$ \\
	& 271.992 & $2s^{2}2p$ $^{2}S_{1/2}$ & $2s2p^{2}$ $^{2}S_{1/2}$ \\
	& 277.264 & $2s^{2}2p$ $^{2}P_{3/2}$ & $2s2p^{2}$ $^{2}S_{1/2}$ \\[1mm]
S X	& 259.497 & $2s^{2}2p^{3}$ $^{4}S_{3/2}$ & $2s2p^{4}$ $^{4}P_{3/2}$ \\
	& 264.231 & $2s^{2}2p^{3}$ $^{4}S_{3/2}$ & $2s2p^{4}$ $^{4}P_{5/2}$ \\[1mm]
Fe X & 174.531 & $3s^{2}3p^{5}$ $^{2}P_{3/2}$ & $3s^{2}3p^{4}3d$ $^{2}D_{5/2}$ \\
       	& 177.240 & $3s^{2}3p^{5}$ $^{2}P_{3/2}$ & $3s^{2}3p^{4}3d$ $^{2}P_{3/2}$ \\
  	& 184.537 & $3s^{2}3p^{5}$ $^{2}P_{3/2}$ & $3s^{2}3p^{4}3d$ $^{2}S_{1/2}$ \\
  	& 207.449 & $3s^{2}3p^{5}$ $^{2}P_{3/2}$ & $3s^{2}3p^{4}3d$ $^{2}F_{5/2}$ \\[1mm]
Fe XI & 180.401 & $3s^{2}3p^{4}$ $^{2}P_{2}$ & $3s^{2}3p^{3}3d$ $^{3}D_{3}$ \\
	& 182.167 & $3s^{2}3p^{4}$ $^{3}P_{1}$ & $3s^{2}3p^{3}3d$ $^{3}D_{2}$ \\
	& 188.216 & $3s^{2}3p^{4}$ $^{3}P_{2}$ & $3s^{2}3p^{3}3d$ $^{3}P_{2}$ \\
	& 192.813 & $3s^{2}3p^{4}$ $^{3}P_{1}$ & $3s^{2}3p^{3}3d$ $^{3}P_{2}$ \\
	& 202.424 & $3s^{2}3p^{4}$ $^{3}P_{2}$ & $3s^{2}3p^{3}3d$ $^{3}P_{2}$ \\[1mm]
Fe XII & 186.854 & $3s^{2}3p^{3}$ $^{2}D_{3/2}$ & $3s^{2}3p^{2}3d$ $^{2}F_{5/2}$ \\
	& 186.887 & $3s^{2}3p^{3}$ $^{2}D_{5/2}$ & $3s^{2}3p^{2}3d$ $^{2}F_{7/2}$ \\
	& 192.394 & $3s^{2}3p^{3}$ $^{4}S_{3/2}$ & $3s^{2}3p^{2}3d$ $^{4}P_{1/2}$ \\
	& 193.509 & $3s^{2}3p^{3}$ $^{4}S_{3/2}$ & $3s^{2}3p^{2}3d$ $^{4}P_{3/2}$ \\
	& 195.119 & $3s^{2}3p^{3}$ $^{4}S_{3/2}$ & $3s^{2}3p^{2}3d$ $^{4}P_{5/2}$ \\[1mm]
Fe XIII & 200.021 & 	$3s^{2}3p^{2}$ $^{3}P_{1}$ & $3s^{2}3p3d$ $^{3}D_{2}$ \\
	& 202.044 & $3s^{2}3p^{2}$ $^{3}P_{0}$ & $3s^{2}3p3d$ $^{3}P_{1}$ \\
	& 251.952 & $3s^{2}3p^{2}$ $^{3}P_{2}$ & $3s3p^{3}$ $^{3}S_{1}$ \\[1mm]
Fe XIV & 211.317 & $3s^{2}3p$ $^{2}P_{1/2}$ & $3s^{2}3d$ $^{2}D_{3/2}$ \\
	& 264.789 & $3s^{2}3p$ $^{2}P_{3/2}$ & $3s3p^{2}$ $^{2}P_{3/2}$ \\
	& 270.521 & $3s^{2}3p$ $^{2}P_{3/2}$ & $3s3p^{2}$ $^{2}P_{1/2}$ \\[1mm]
\hline \hline \\[-3mm]
\end{tabular}
\caption{List of EIS EUV Lines Used for EM Loci Temperature Analysis}
\label{tab:EISEMLociLines}
\end{center}
\end{table}
We first coalign the vertical axis at the central wavelength of each emission line to the 186.8 \AA{} FOV, accounting for the wavelength-dependent vertical drift of the EIS detector readout. We then calculate the $EML^{*}$ function for each of these seven ion species and search for an intersection point common to all of them, doing so for spectra sampled at each detector row and raster step. The search algorithm we use to find these intersections operates by pairing every possible combination of $EML^{*}$ curves and calculating secant lines for both curves in every pair.  A single secant is anchored by two point on its associated $EML^{*}$ function, each of which are evaluated at temperatures separated by $\Delta{}\text{log}_{10}T\left[K\right]=0.003$.  The $EML^{*}$ pair is considered to be intersecting if the secants intersect in a temperature range bounded by the anchor points.  In these instances, the recorded temperature would be that associated with the secant intersection point.

With a complete catalogue of intersections for each set of $EML^{*}$ curves at each detector row and raster position, we now determine whether a genuine common intersection exists for each of those sets.  Since the catalogue of intersection points also includes many incidental intersections unrelated to the isothermal nature of the plasma, it is necessary to perform additional processing to eliminate those points.  To do so, we apply a clustering algorithm to each $EML^{*}$ set.  First, each coordinate axis is transformed from a logarithmic to a linear scale and then normalized to range from 0 to 1, corresponding to the minimum and maximum values, respectively, of $T$ and $EML^{*}\left(T\right)$.  The algorithm then focuses on the leftmost intersection point which we call the \textit{target point}.  The normalized distances from the target point to every other point is then calculated.  Next,  we fill the coordinate space with concentric rings of equal width centered on the target point.  The ring width is initially set to the equivalent normalized length of $T=200$ K on the x-axis.  If at least seven other intersection points lie within the central ring, then the target point is considered to be part of a cluster.  If not, then we gradually increase the ring width by increments of 200 K until one of the two following conditions are met: (1) the ring size exceeds 200,000 K, or (2) at least seven points lie within one ring. If condition (1) occurs then the target point is deemed not to be part of a cluster.  If condition (2) occurs, then we check that the points lie within the central ring.  If not, then we consider the target point to be not part of a cluster; if so, then we consider the target point to be part of a cluster.  We then repeat this process for all intersection points. We then check for multiple clusters by ensuring that no two intersection points identified as clustered are farther than 300,000 K apart.  This leaves us with the following three ending scenarios: (1) there is a single cluster of intersection points, (2) there are multiple clusters of intersection points, or (3) there are no clusters of intersection points.  For scenario (1), we consider the characteristic isothermal temperature of the plasma to be the mean temperature of the cluster points.  For scenarios (2) and (3), we consider the EM loci method to have failed and do not record a temperature for that location.

The AIR-Spec data require a modified approach.  Unlike the EIS data, the AIR-Spec data do not benefit from having a large number of detected lines. Each of the four lines corresponds to a unique ion species, so we do not use the modified EM loci function, $EML^{*}$, applied to the EIS data. Instead, we slightly modify Eq. (\ref{eq:eml}) by replacing the contribution function, $G_{ij}$, with the volume emissivity, $\epsilon_{ij}$.  This allows us to incorporate radiative processes that may influence the IR lines but not the far more resilient EUV lines.  Also, the AIR-Spec data do not benefit from the strong signal-to-noise ratio that the EIS data has.  So, instead of applying the EM loci analysis to each detector row and raster position, we instead coadd the spectra across all exposures and take samples at 30 arcsec and 100 arcsec from the limb.

\section{Results and Discussion}\label{sec:RandD}
\subsection{Electron Densities}

\begin{figure}[!hbtp]
\begin{center}
\includegraphics[trim=4cm 0cm 4cm 2cm,clip=true,width=\linewidth]{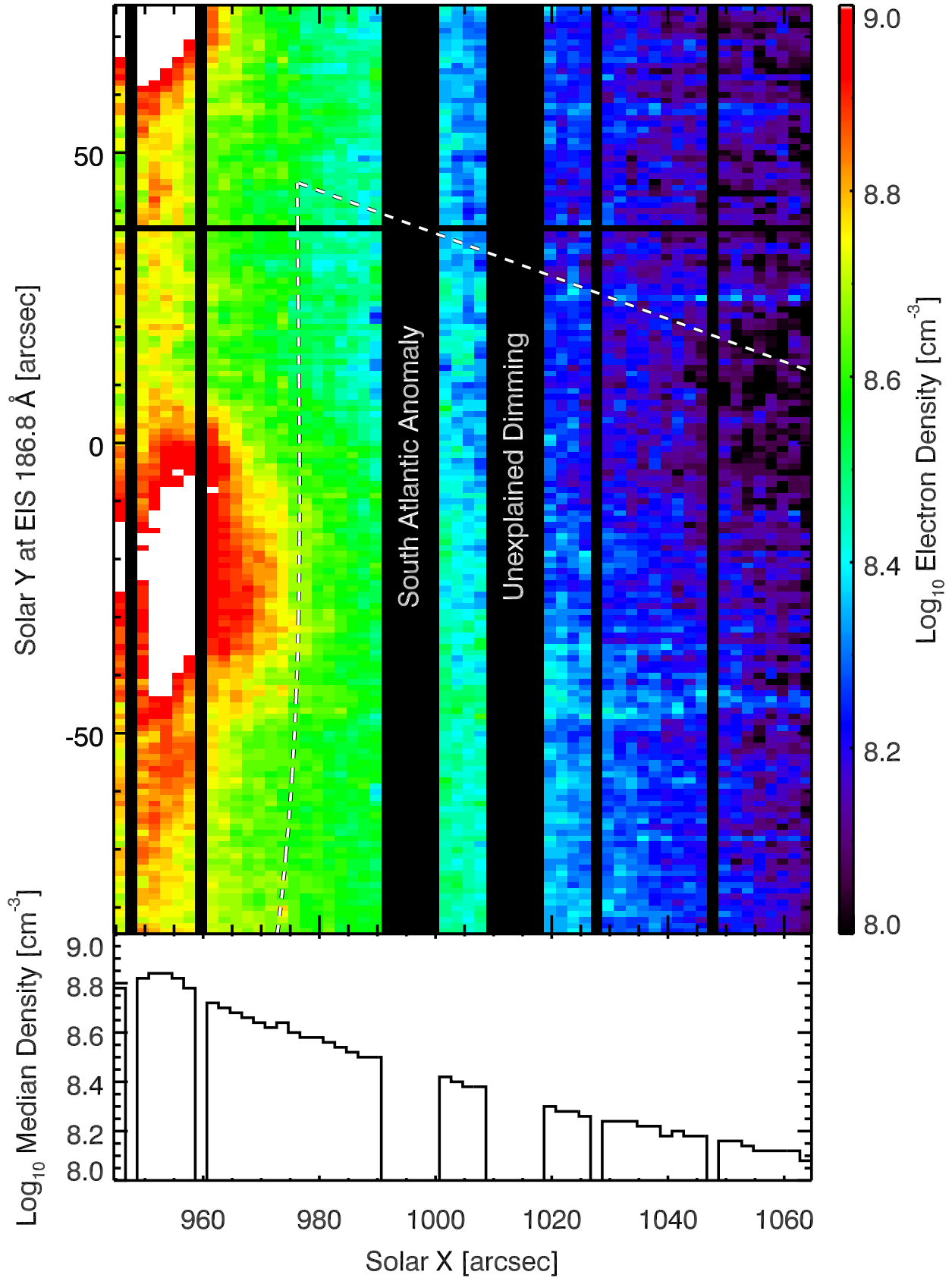}
\caption{Map of line-of-sight integrated electron densities throughout the EIS field-of-view (top panel.) Black vertical and horizontal bars indicate data not included in the analysis due to the South Atlantic anomaly, the unexplained dimming event, frame dropouts, or other detector defects.  The dashed line indicates the northeastern boundary of the AIR-Spec field-of-view.  The lower panel depicts the columnar median of electron density values across the solar X dimension of the EIS field-of-view.}
\label{fig:densemap}
\end{center}
\end{figure}

As seen in Figure \ref{fig:densemap}, the strong signal-to-noise ratio of Fe XII 186.85/.89 \AA{} and Fe XII 192.39 \AA{} throughout the observation allowed us to produce a detailed map of the coronal electron density resolved to instrumental limits.
This map depicts electron densities spanning an order of magnitude from $10^{8}$ cm$^{\text{-}3}$ at the western extreme of the EIS FOV to $10^{9}$ cm$^{\text{-}3}$ in localized regions close to the limb. 

\begin{figure}[!hbtp]
\begin{center}
\includegraphics[width=0.5\linewidth, angle=90]{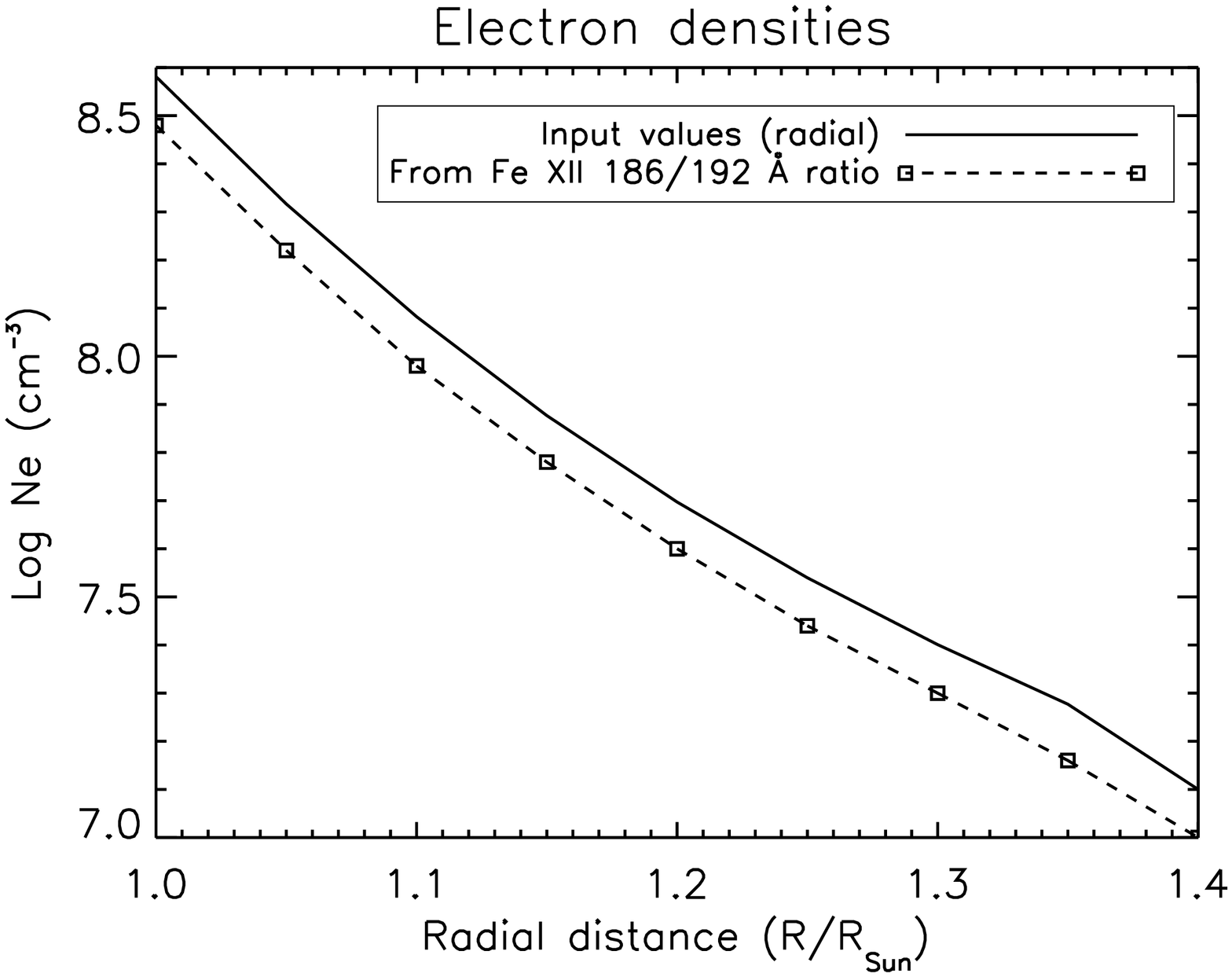}
\caption{The assumed variation of the radial electron density (solid line) from UVCS inputs, with the 
expected values one would obtain from the Fe XII ratio derived from CHIANTI for each location off the limb.}
\label{fig:ne_profile}
\end{center}
\end{figure}

We find a strong correlation between the electron density structure of the coronal plasma and the morphological features depicted in the AIA images of the same region seen in Figure \ref{fig:AIAcutcon}.  Two localized elliptical regions containing the highest densities just off the limb \textemdash{} one spanning from $Y=\text{-}50$ arcsec to 0 arcsec, and the other spanning from $Y=50$ arcsec to some location north of the FOV \textemdash{} are cospatial with intensity enhancements seen in AIA 193 \AA{} and 211 \AA{}, which overlap the wavelength range of EIS, and to a lesser extent 94 \AA{} and 335 \AA{}.  The enhancements appear to be the setting remnants of small-scale flux emergence unassociated with an active region.  Although they are near the likely footpoints of the prominence, the enhancements are unlikely to be related since the prominence does not appear to undergo any activation or eruption during or after the EIS observations. Additionally, at the far western end of the map centered near $Y=0$ arcsec is a semi-circular region of excess density depletion.  This is cospatial with eastern end of a large coronal cavity seen in AIA 193 \AA{} which formed coincident to the prominence. 

The electron density decays remarkably smoothly from the limb outward. The bottom panel of  Figure \ref{fig:densemap} shows the median electron density in each column of the map.  With the exception of the small sliver east of $X=950$ occupied by the solar disk, and despite the South Atlantic anomaly, unexplained dimming phenomenon, and numerous frame dropouts, this plot shows a clear exponential decay of electron density consistent with other quiet-Sun measurements.  In particular, Figure \ref{fig:ne_profile} shows radial semi-empirical models of the electron density under quiet-Sun conditions derived from CHIANTI predictions and data from the Ultraviolet Coronagraph Spectrometer \citep[UVCS,][]{Kohl1995} onboard SOHO, similar to the models detailed in \citet{DelZanna2018c}. At 1.1 R$_{\odot}$, corresponding approximately to the western end of the EIS FOV, Figure \ref{fig:ne_profile} shows modeled electron densities between $10^{8.1}$ cm$^{-3}$ and $10^{8.2}$ cm$^{-3}$, agreeing strongly with the results depicted in Figure \ref{fig:densemap}. The prominence's presence does not appear to alter this quiet-Sun trend despite likely retaining much of its partially ionized plasma from the chromosphere.  This is likely a consequence of the prominence's limited spatial extent and low opacity across much of the EUV which allows the hot, quiescent, fully ionized coronal plasma to dominate the line-of-sight source volume. Overall, the electron density measurements appear to be reliable and will provide a suitable supporting role in the EM loci analysis.

\subsection{Isothermal Temperatures and IR Diagnostic Potential}

\begin{figure}[!hbtp]
\begin{center}
\includegraphics[trim=2cm 0cm 0cm 0cm,clip=true,width=\linewidth]{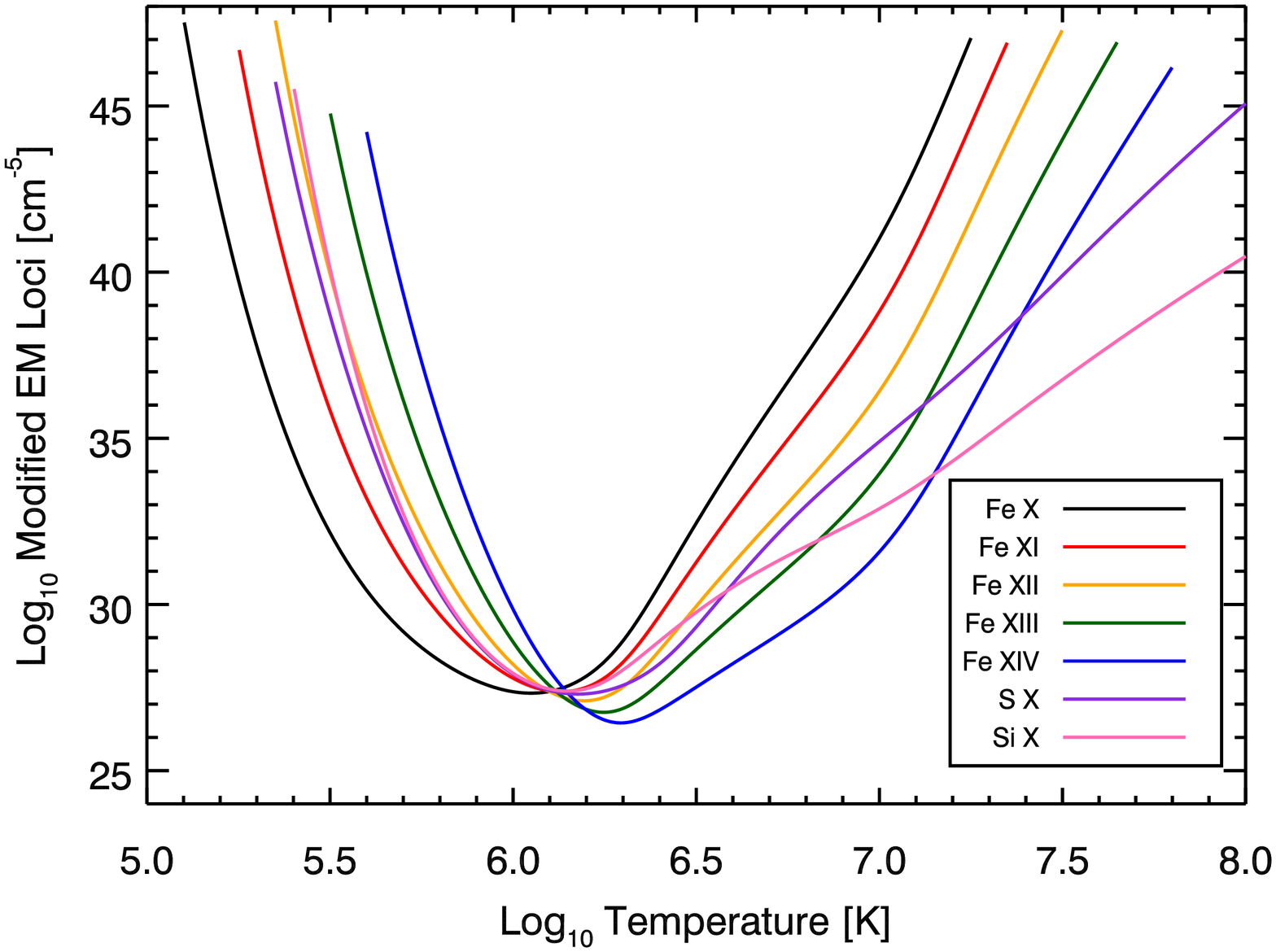}
\caption{Example of modified EM loci ($EML^{*}$) functions calculated for seven ions associated with emission lines detected by EIS.  This example was constructed from data sampled near solar $\left(X,Y\right)=\left(966,11\right)$ arcsec. The $EML^{*}$ curves show a strong tendency to intersect one another near $T=10^{6.1}$ K, indicating an isothermal plasma characterized by that temperature. Note that uncertainties are too small to meaningfully depict on the plot.}
\label{fig:EISEM}
\end{center}
\end{figure}

\begin{figure}[!hbtp]
\begin{center}
\includegraphics[trim=4cm 0cm 4cm 2cm,clip=true,width=\linewidth,hiresbb=true]{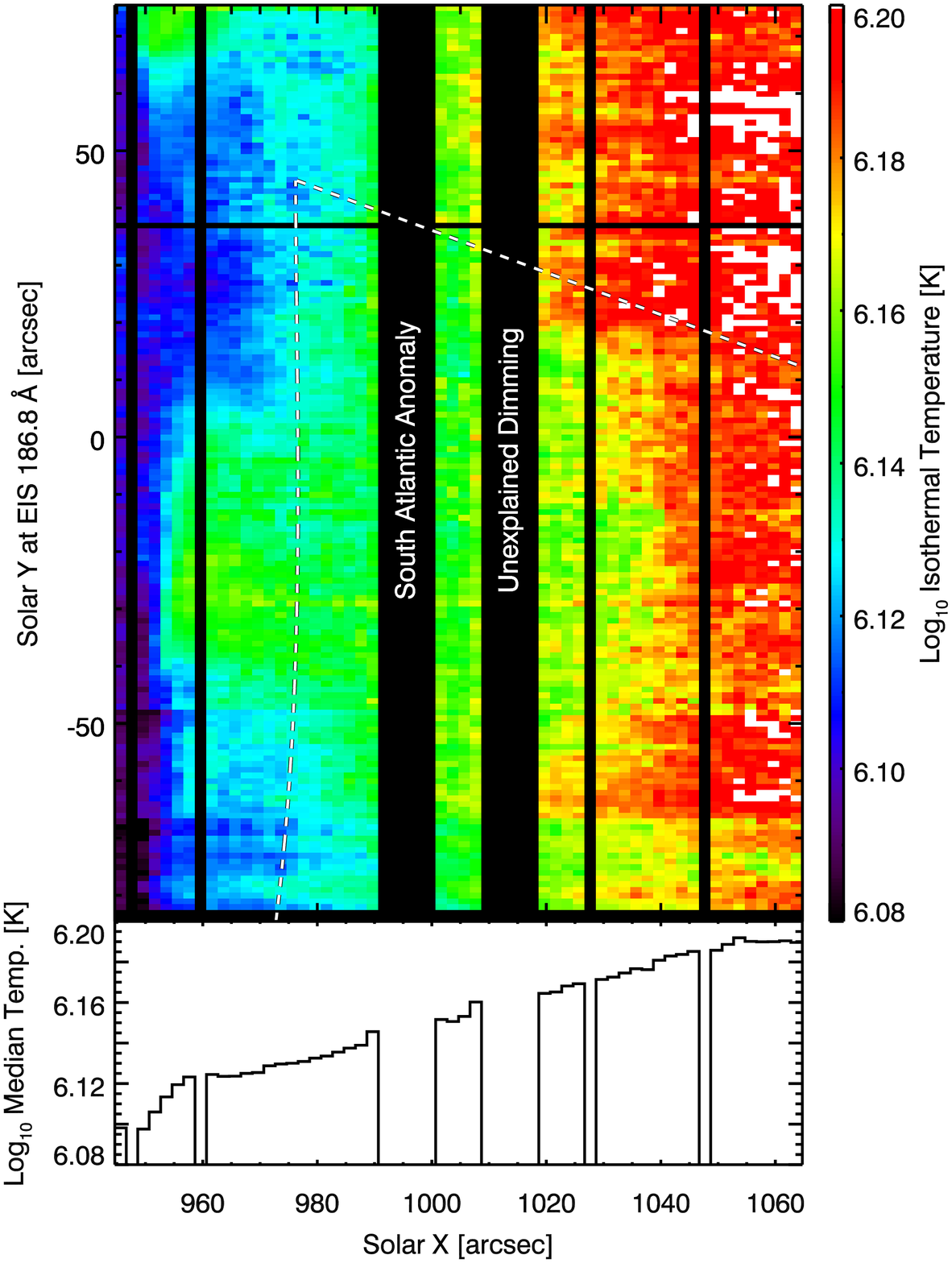}
\caption{Map of isothermal temperatures derived from EM loci analysis in the EIS field-of-view (top panel.) Black vertical and horizontal bars indicate data not included in the analysis due to the South Atlantic anomaly, the unexplained dimming event, frame dropouts, other detector defects, or failure of the $EML^{*}$ curves to form a precise intersection point. The dashed line indicates the northeastern boundary of the AIR-Spec field-of-view. The lower panel depicts the columnar median of temperature values across the solar X dimension of the EIS field-of-view.}
\label{fig:tempmap}
\end{center}
\end{figure}

The EM loci temperature diagnostics were successful across a vast majority of the EIS data set.  All sampled locations in the FOV produced a clustered set of intersection points with the exception of two detector rows at the bottom of the FOV where there were insufficient data from long-wavelength windows due to the wavelength-dependent vertical shift of the EIS spectra. Figure \ref{fig:EISEM} depicts a typical EM loci plot for a single sample within the EIS data set.
It demonstrates a tightly bound cluster of intersection points within a temperature range of $10^{6.1}$ to $10^{6.2}$, a feature we found to be ubiquitous across all but a small few samples. As a result, we can produce a similarly well-resolved map of the isothermal temperature, seen in Figure \ref{fig:tempmap}, as we did for the electron density analysis.
The map shows temperatures spanning from $10^{6.1}$ K near the limb to $10^{6.2}$ K near the western edge of the EIS FOV.  As is the case with the electron density map, the isothermal temperature exhibits a smooth, exponential trend as seen in the bottom panel of Figure \ref{fig:tempmap}; however, unlike the electron density, the isothermal temperature generally increases with distance from limb as expected for quiescent lower coronal plasma.  

We also see a correlation between the localized intensity enhancements seen in AIA 94 \AA{}, 193 \AA{}, 211 \AA{} and 335 \AA{} \textemdash{} four passbands which tend to image hotter plasma than the others \textemdash{} and temperature enhancements on the order of $\Delta{}\text{log}_{10}T\left[K\right]=0.04$. It is unclear whether the prominence influenced the temperature diagnostics.  Other than the temperature enhancements near the limb, the only anomalous temperature deviation appears in a region bounded by solar $X=1020\to{}1040$ arcsec and solar $Y=\text{-}50\to{}20$ arcsec.  This region is characterized by an intrusion of cooler plasma on the order of $10^{6.16}$ K into the $10^{6.20}$ K plasma near the western edge of the FOV.  This is cospatial with the farthest reaches of the main prominence arch; however, since there is no apparent signature of the prominence in the electron density structure, it's unclear whether it is related to this temperature feature.  Furthermore, the region is bounded from below by what appears to be a horizontal discontinuity.  This represents the southern extent of the FOV for the first long-wavelength window of the full-CCD readout.  Below this point, the number of available species for EM loci analysis drops, which appears to have some systematic effect of the temperature measurements.  Given this, it's difficult to say whether the lower-temperature structure is correlated with the contour of the prominence arch or simply a continuation of a larger lower-temperature structure originating from below the FOV.  Overall, the results of the EM loci analysis on the EIS data appear reliable and can serve as a good benchmark for the AIR-Spec temperature measurements.

\begin{figure}[!hbtp]
\begin{center}
\includegraphics[width=0.8\linewidth]{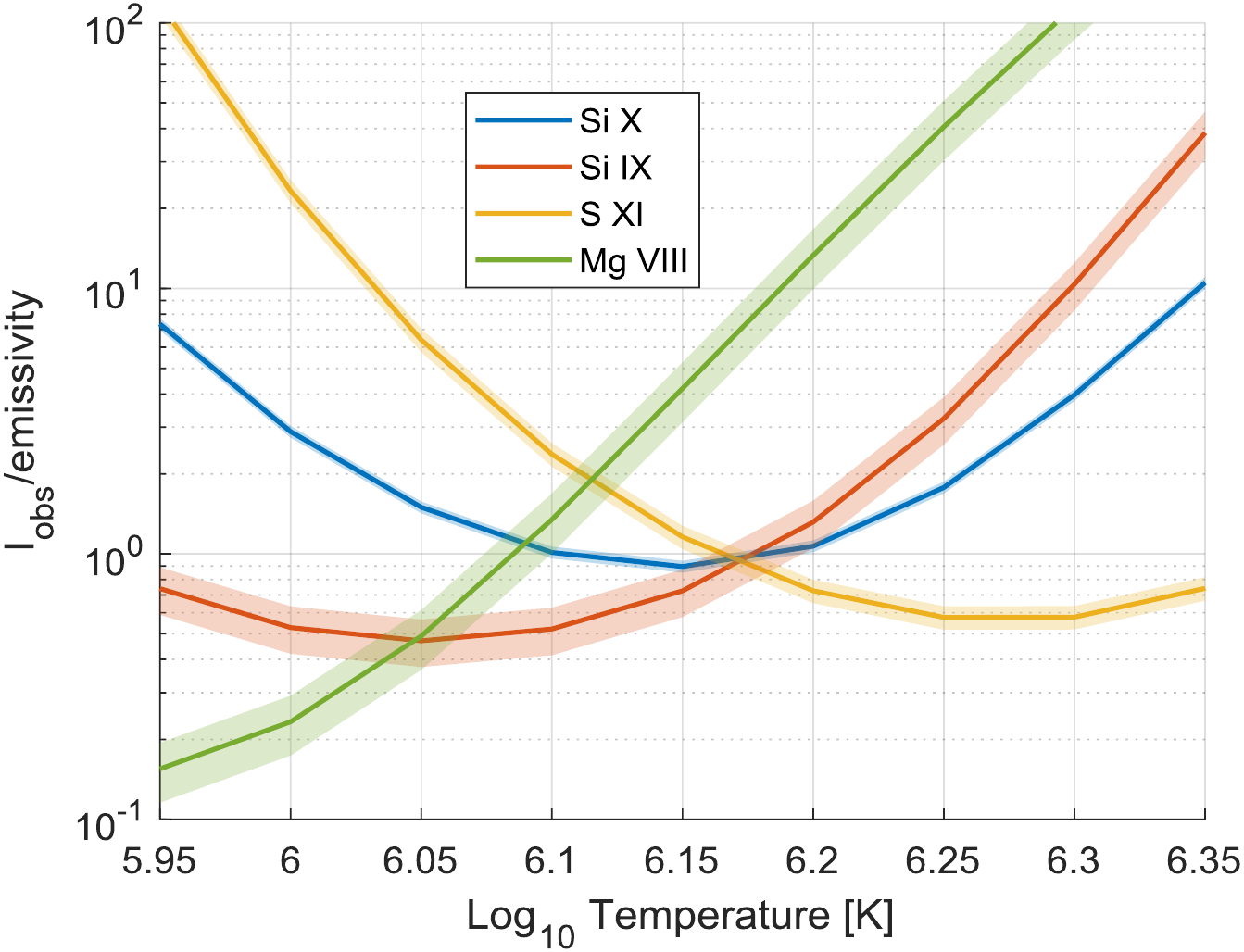}
\caption{Plot of observed AIR-Spec IR line intensities sampled at 30 arcsec from the limb divided by the volume emissivity, which we use as a proxy for the EM loci function. The shaded envelopes around the curves depict the 95\%{} confidence interval due to uncertainties arising from Gaussian fits to the line profiles. The intersection points are more spread out than those seen in the EIS data, but are still compact enough to show isothermal temperatures in the range of $10^{6.0}$ to $10^{6.2}$ K.}
\label{fig:AIRSpecEM30}
\end{center}
\end{figure}

\begin{figure}[!hbtp]
\begin{center}
\includegraphics[width=0.8\linewidth]{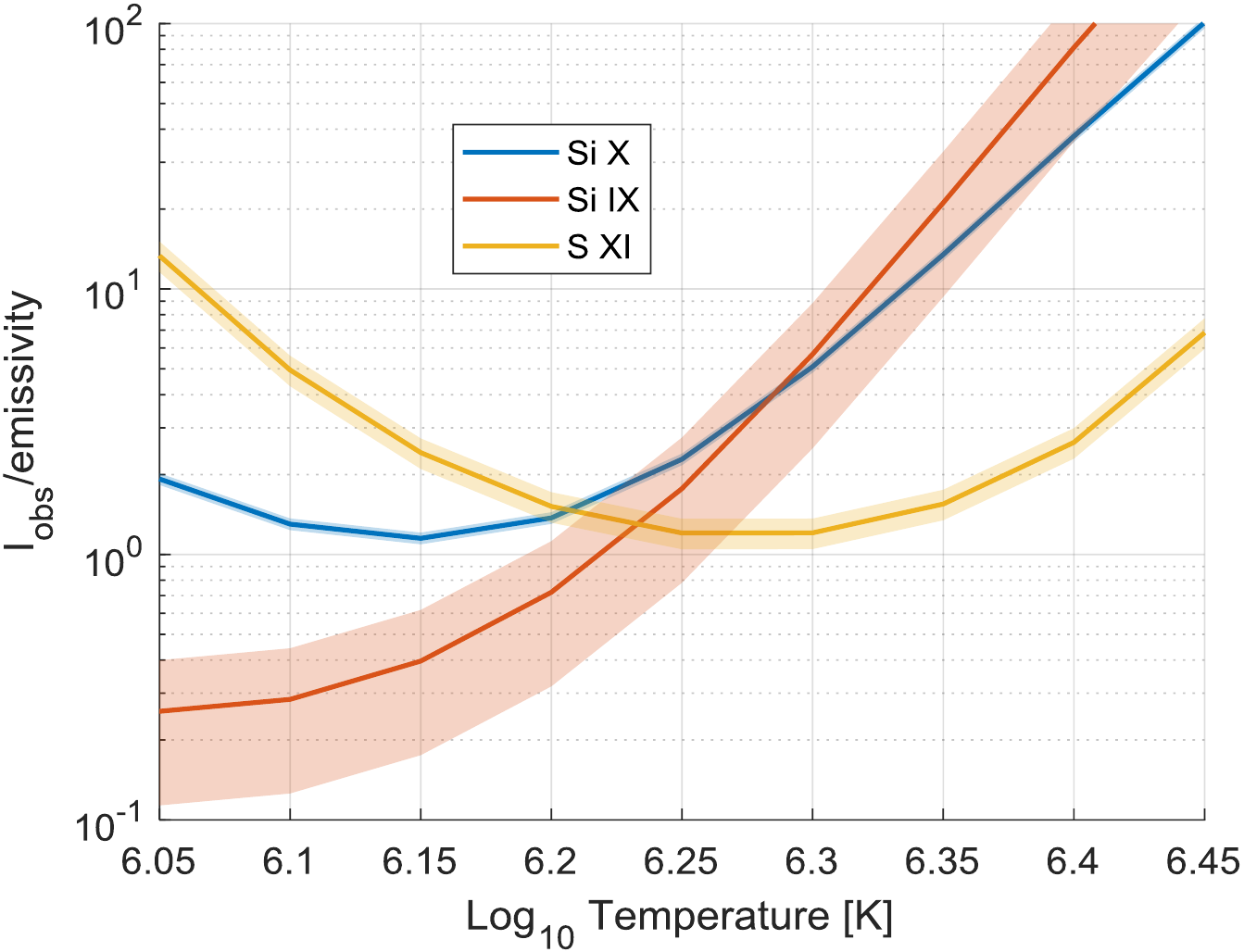}
\caption{Plot of observed AIR-Spec IR line intensities sampled at 100 arcsec from the limb divided by the volume emissivity, which we use as a proxy for the EM loci function. The shaded envelopes around the curves depict the 95\%{} confidence interval due to uncertainties arising from Gaussian fits to the line profiles. The weakest line, Mg VIII 3.03 \textmu{}m, was not detected at this location. Although the number of intersection points is small, there is a noticeable shift towards higher temperatures than depicted in Figure \ref{fig:AIRSpecEM30}.}
\label{fig:AIRSpecEM100}
\end{center}
\end{figure}

The EM loci analysis of the AIR-Spec data, although not as robust, is consistent with the EIS results.  Figures \ref{fig:AIRSpecEM30} and \ref{fig:AIRSpecEM100} depict the results of the EM loci analysis for AIR-Spec data sampled at 30 arcsec and 100 arcsec from the limb, respectively.  The shaded bands indicate 95\% confidence intervals due to uncertainty in the Gaussian fits to each line. Errors in the AIR-Spec radiometric calibration and modeled volume emissivity are not indicated, but likely contribute to the spread in intersection points of the EM loci curves.


Both samples display a spread of intersection points broader than what is typically seen in the EIS data, and neither sample would pass the strict clustering standards established for the EIS data in Subsection \ref{subsec:tempdiag}.  However, this does not wholly invalidate the results of the AIR-Spec temperature diagnostics, since both show central tendencies consistent with EIS temperature measurements.  For the 30-arcsec sample, the ten EM loci intersections span a range of temperatures from $10^{6.05}$ K to $10^{6.17}$ K with a variance-weighted mean temperature of $10^{6.13\pm{}0.02}$ K. For the 100-arcsec sample, we only have three intersections since Mg VIII 3.03 \textmu{}m was not detected at this distance from the limb. The intersections span a range of warmer temperatures from $10^{6.20}$ K to $10^{6.28}$ K with a variance-weighted mean temperature of $10^{6.21\pm{}0.01}$ K.  Both mean temperatures are consistent with their EIS counterparts, capturing the overall outwardly increasing trend of the quiescent coronal plasma temperature. Note that the reported uncertainties are dispersion-corrected standard deviations of the variance-weighted means.  When scaled logarithmically, the uncertainty spits into unequal upper and lower components; however, since these components do not differ by more that 0.01 in both cases, we simply quote a single uncertainty down to that precision.

\begin{figure}[!hbtp]
\begin{center}
\includegraphics[width=0.5\linewidth, angle=90]{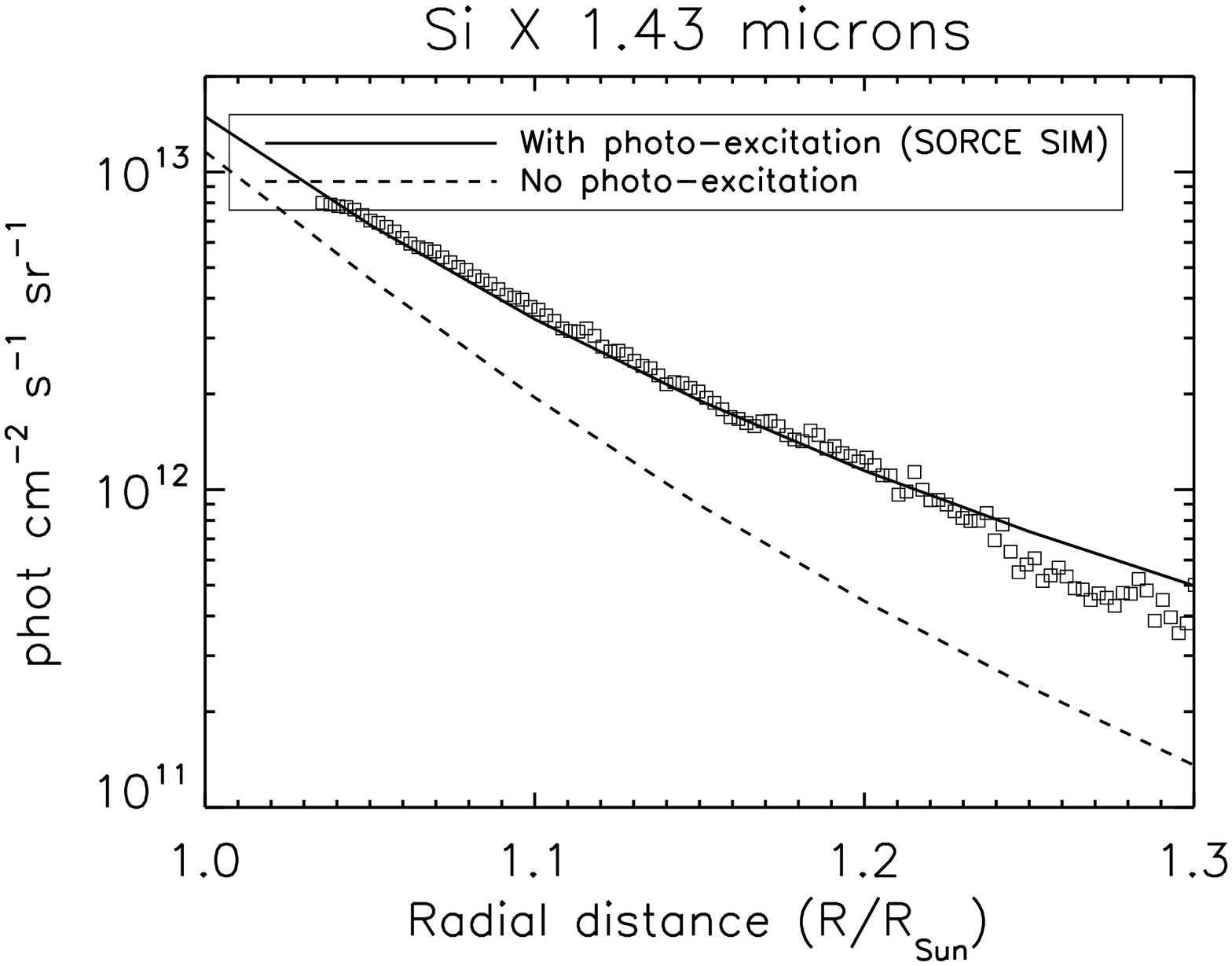}
\caption{Predicted radial decrease of the strongest AIR-Spec line, Si X 1.43 \textmu{}m, with photo-excitation form SORCE TIM 
(solid line) and without photo-excitation (dashed line). The observed AIR-Spec radiances, scaled, are shown
with boxes.}
\label{fig:si_x}
\end{center}
\end{figure}

We must address the breakdown of physical assumptions that are valid in the EUV but not in the IR.  Of critical interest is the effect of photoexcitation from solar IR continuum.  We attempted to correct for this by using the volume emissivity, $\epsilon_{ij}$, in place of the contribution function, $G_{ij}$, for the IR EM loci analysis. This allowed us to account for photoexcited effects on level populations. Figure \ref{fig:si_x} shows the radial change in radiance for the Si X 1.43 \textmu{}m observed by AIR-Spec and two predictions of the same value derived from CHIANTI. One prediction accounts for photoexcitation using continuum irradiance measurements from the Total Irradiance Monitor \citep[TIM,][]{Kopp2005} onboard \textit{Solar Radiation and Climate Experiment} (SORCE) while the other does not.  The radial decay of AIR-Spec Si X 1.43 \textmu{}m radiances agrees strongly with the photoexcited model within the EIS FOV (i.e. within 1.1 R$_{\odot}$.) However, the AIR-Spec Si X 1.43 \textmu{}m are scaled to overlap the photoexcited model in Figure \ref{fig:si_x}, so the observed data only agrees with the rate of radial decay and not the predicted radiances themselves.  Although the agreement is promising, it does not rule out dramatic systematic changes of the near-limb AIR-Spec Si X 1.43 \textmu{}m radiances due to local continuum enhancements or variations due to the solar cycle.  

The cool plasma from the prominence also introduces uncertainty into the analysis.  The plasma trapped in the prominence likely originates from the upper chromosphere where we typically find partially ionized plasma at temperatures approaching $10^{5}$ K.  When injected into the new collisional and radiative environments of the corona, the prominence plasma is unlikely to maintain ionization equilibrium. If the prominence proves to be a significant source of coronal IR emission, then the assumptions surrounding our electron density measurements may fail.  However, the prominence may actually be represented in the EM loci analysis of the AIR-Spec data. Figure \ref{fig:AIRSpecEM30} features Mg VIII 3.03 \textmu{}m, a line that forms at cooler temperatures than the other AIR-Spec lines and is only seen near the limb.  The EM loci curve associated with this line strays far from the common intersection point of the other three EM loci functions ($\sim{}10^{6.17}$ K), trending towards cooler temperatures.  It is possible that the this is an indication of a two-component plasma, a quiet-Sun component that occupies the entire AIR-Spec FOV and a cooler prominence component only observed near the limb.  However, there is considerable uncertainty surrounding this claim as we only provide one instance of this phenomenon and more EM loci curves associated with cooler lines are needed to distinguish between two distinct intersection points and a single intersection point encumbered by additional sources of uncertainty.

Finally, it's important to address the role atomic abundances play in the EM loci analysis.  Precise intersections will only form under isothermal conditions if we assume the correct abundances for all atomic species involved.  In this work, we assume photospheric abundances from \citet{Asplund2009}, which we find well approximate the coronal abundances for the four atomic species that take part in the analysis: Mg, Si, S, and Fe. \citet{DelZanna2018c} found abundances for Mg and Si in the near-limb, quiet-Sun corona close to their respective photospheric values after reprocessing SUMER and UVCS data.  This is consistent with the observed first ionization potential (FIP) bias for coronal abundances where elements with high FIP, such as O (13.6 eV), are far more underrepresented in the corona when compared to their photospheric values than elements of relatively low FIP, such as Mg (7.6 eV) and Si (8.2 eV) \citep[e.g.][]{Raymond1997}. According to this effect, we also expect Fe (7.9 eV) to be near photospheric abundance in the corona, with the possibility of a slight enhancement \citep[e.g.][]{Aschwanden2003,Phillips2012}. However, it's unclear how the FIP bias affects the coronal abundance of an element with intermediate FIP, such as S (10.4 eV). In this case, we look to the results of our EM loci analysis as applied to the AIR-Spec lines. As seen in Figures \ref{fig:AIRSpecEM30} and \ref{fig:AIRSpecEM100}, the EM loci curve for S XI precisely intersects the curves of both Si IX and Si X at a common point, implying that the coronal abundance of S must also be near its photospheric value. 

\section{Conclusions}\label{sec:concl}

In this work, we used coordinated EUV and IR spectral observations of the August 21, 2017 total solar eclipse to characterize coronal plasma near the limb.  Full-CCD EIS observations provided robust diagnostics for electron densities.  This made EM loci temperature diagnostics possible for both the EIS and AIR-Spec data sets.  We found consistent temperature measurements for both data sets, featuring an outward increase from about about $10^{6.1}$ K near the limb to about $10^{6.2}$ K around 100 arcsec from the limb. We conclude that IR emission lines have great potential to act as temperature diagnostics for coronal plasma when coupled with electron density measurements from coordinated EUV spectral observations.

However, this conclusion is not without some uncertainty, especially considering the difference in quality between the EIS and AIR-Spec data sets.  We expect to remedy this during the second flight of AIR-Spec on July 2, 2019 to observe a total solar eclipse over the southern Pacific Ocean.  The 2019 observation will feature increased sensitivity and reduced jitter.  Improvements to thermal shielding in the IR camera and spectrometer will reduce the background level by a factor of 15 to 25, improving the signal-to-noise ratio by a factor of 4 to 5. Closing the loop on the image stabilization system will reduce the jitter by a factor of 20, allowing us to better coordinate with other observatories by targeting precise locations in the corona or scanning the slit systematically.

Our results bear considerable importance with regards to anticipated advancements in ground-based IR observations of the Sun.  Of particular interest is DKIST, which is under construction at the Haleakala Observatory in Hawaii, USA. This observatory will boast a state-of-the-art suite of visible and infrared instruments, including the Cryogenic Near-infrared Spectro-polarimeter \citep[Cryo-NIRSP,][]{Fehlmann2016} and the Diffraction-limited Near-infrared Spectro-polarimeter \citep[DL-NIRSP,][]{Elmore2014}, capable of observing at least three of the four AIR-Spec target lines (Si X 1.431 \textmu{}m, Mg VIII 3.03 \textmu{}m, and Si IX 3.94 \textmu{}m) as well as other IR coronal lines with potential diagnostic significance (e.g. Fe XIII 1.07 \textmu{}m, Fe IX 2.22 \textmu{}m, and Si IX 2.58 \textmu{}m.)  Coordinated EUV spectral observations will prove to be invaluable in both evaluating the diagnostic value of DKIST's observed IR lines, and providing critical supplementary information needed for a complete physical characterization of coronal plasma.

\acknowledgements{}

\textit{Hinode} is a Japanese mission developed and launched by ISAS/JAXA, collaborating with NAOJ as a domestic partner, and NASA and UKSA as international partners. Scientific operation of the \textit{Hinode} mission is conducted by the \textit{Hinode} science team organized at ISAS/JAXA. This team mainly consists of scientists from institutes in the partner countries. Support for the post-launch operation is provided by JAXA and NAOJ (Japan), UKSA (U.K.), NASA, ESA, and NSC (Norway). EUV context imaging is courtesy of NASA/\textit{SDO} and the AIA science team. AIR-Spec was developed under an NSF Major Research Instrumentation grant, AGS-MRI 1531549, with cost sharing by The Smithsonian Institution.

\bibliographystyle{aasjournal}
\bibliography{EIS_AirSpec_Paper.bbl}

\end{document}